\documentclass[twocolumn,amsmath]{astronomycomm}

\makeatletter
\renewcommand{\citep}[1]{(\ac@citeloop#1,\@nil)}
\def\ac@citeloop#1,#2\@nil{%
  \citeauthor{#1}, \citeyear{#1}%
  \def\ac@rest{#2}%
  \ifx\ac@rest\@empty\else; \ac@citeloop#2\@nil\fi}
\makeatother

\newcommand{\OmF}{\Omega_{F}}
\newcommand{\OmH}{\Omega_{H}}

\shorttitle{An exact engine for black-hole jets}
\shortauthors{Wang}

\begin{document}

\title{An Exact Engine for Black-Hole Jets: A Self-Gravitating Einstein--Maxwell Split-Monopole Solution}

\author[orcid=0000-0001-7959-3387,gname=Yu,sname=Wang]{Yu Wang}
\affiliation{ICRA, Dipartimento di Fisica, Sapienza Universit\`a di Roma,
Piazzale Aldo Moro 5, I-00185 Roma, Italy}
\affiliation{ICRANet, Piazza della Repubblica 10, I-65122 Pescara, Italy}
\affiliation{INAF -- Osservatorio Astronomico d'Abruzzo, Via M. Maggini
snc, I-64100 Teramo, Italy}
\affiliation{Marcel Grossmann Center, Brickell Avenue 701, Miami,
FL 33131, USA}
\email[show]{yu.wang@icranet.org}

\begin{abstract}
For fifty years the Blandford--Znajek mechanism has been a mechanism and not an exact solution: force-free electrodynamics on a prescribed metric, with the jet's own field carrying no weight. Here that field gravitates. A split monopole weighs as much as a magnetic monopole, so its self-gravity is the magnetic Reissner--Nordstr\"om geometry; two copies glued across an equatorial current sheet put hole, disk and jet-driving flux into one spacetime, and force-free plasma makes a magnetosphere of it. For this idealized configuration the static solution is exact at any magnetic flux, the rotating one is exact at first order in rotation and at any flux, and extraction is treated adiabatically. Three things then follow that no test-field calculation can see. First, a steady jet is impossible. A stationary horizon cannot be heated but a slipping magnetosphere necessarily heats it, so the only stationary state is a dead one corotating with the hole, and a working jet is a black hole in decay at rates the field equations fix rather than assume. Second, flux enters the laws of black-hole mechanics as a charge, sharing one extremality budget with spin. That budget caps the jet power at $c^5/48G$, at leading order in rotation, whatever the mass. Third, the lifetime output is finite: a maximally spinning hole delivers $1-e^{1/4}/\sqrt2=9.2\%$ of its mass and grows its horizon area by exactly $\sqrt e$. In the maximal flux limit the horizon's own moment of inertia vanishes with the irrational exponent $(\sqrt{17}-1)/2$ set by the near-horizon throat, and what is left is a rotating hole whose angular momentum has passed to its own field.
\end{abstract}


\section{Introduction}

Relativistic jets from accreting black holes are understood through the mechanism of \citet{1977MNRAS.179..433B}, hereafter BZ, a continuation and development of studies of black-hole energy extraction, magnetized accretion and pulsar electrodynamics \citep{1969NCimR...1..252P,1969ApJ...157..869G,1969Natur.223..690L,1970PhRvL..25.1596C,1973CMaPh..31..161B,1974ApJ...193..443T,1974PhRvD..10.1680W,1975PhRvD..12.2959R}. Magnetic field lines thread the horizon, the hole's rotation twists them, and the twist carries energy and angular momentum away as a Poynting flux. The picture underlies the interpretation of the M87 image \citep{2019ApJ...875L...1E}, the magnetically arrested states of simulations \citep{2003PASJ...55L..69N,2011MNRAS.418L..79T}, and analytic jet-power estimates.

It has never been an exact solution. What exists is force-free electrodynamics on a prescribed metric, usually Kerr, with the jet's own field treated as a test field: the field feels the geometry and the geometry never feels the field. That covers the original perturbative calculation and its modern high-order refinements \citep{2008PhRvD..78b4004T,2015PhRvD..91f4067P,2018PhRvD..98h4056G,2020JCAP...04..009A,2022JCAP...07..032C}, its extension to nonlinear electrodynamics \citep{2023AnP...53500635C}, the membrane paradigm's horizon thermodynamics \citep{1977MNRAS.179..457Z,1978PhRvD..18.3598D,1986bhmp.book.....T}, and the causality debate around the steady state \citep{1990ApJ...350..518P,2004MNRAS.350..427K}. Even the one study of metric response computed the linear reaction to a prescribed field \citep{2021PTEP.2021i3E03K}.

We put the field into the Einstein equations and solve the coupled system, which gives its reaction back on the hole and the limits the whole system sets on the jet; the solution is exact in the sense that its equations and boundary conditions hold identically, for an idealized configuration with an equatorial current sheet in place of a disk. Section~\ref{sec:spacetime} builds that spacetime, Section~\ref{sec:steady} shows that the engine it carries cannot be steady, Section~\ref{sec:budget} charges the flux to the hole's extremality budget and reads off the ceilings that follow, and Section~\ref{sec:endpoint} examines the maximal flux limit. Section~\ref{sec:scope} separates what is proved from what is computed, and Section~\ref{sec:conclusions} concludes.\footnote{The main text carries the physical picture and the results, in a form readers from any field can follow quickly; the appendices carry the technical calculations in enough detail to check and reproduce them. The symbolic and numerical routines are at \url{https://github.com/YWangScience/An-Exact-Engine-for-Black-Hole-Jets}.}

\section{The engine as a spacetime}\label{sec:spacetime}

\begin{figure}
\centering
\includegraphics[width=0.99\linewidth]{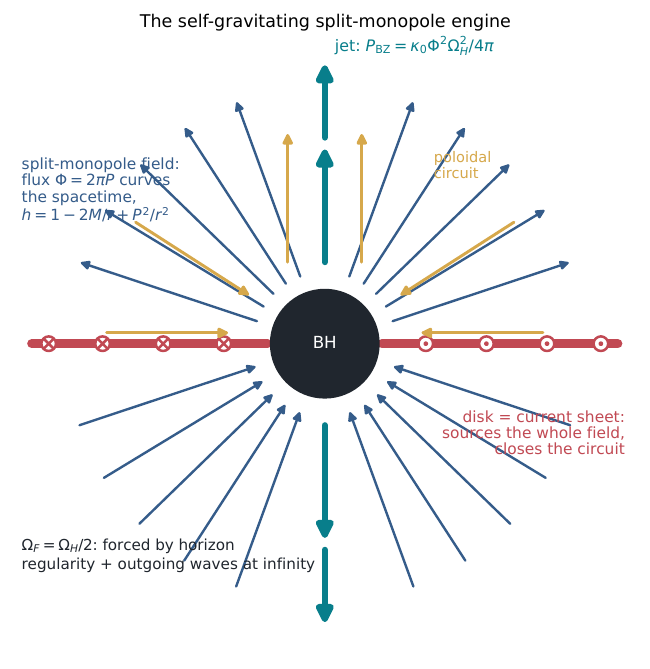}
\caption{The solution. An equatorial sheet of azimuthal current, the minimal disk, sources a split-monopole field: outward in the north, inward in the south, flux $\Phi=2\pi P$ through each horizon hemisphere. Because the field's stress is blind to its sign, the spacetime is exactly magnetic Reissner--Nordstr\"om on both sides of the sheet, with the flux sitting in the metric. When the hole spins, a poloidal current loop closes through the sheet and Poynting flux leaves along the field lines. No part of the configuration is treated as a test field.}
\label{fig:anatomy}
\end{figure}

One field configuration makes the exact treatment possible, and it is the one BZ theory already uses. Magnetic stress-energy is quadratic in the field, so reversing the field across the equator leaves the stress untouched. A split monopole does exactly that, its field running radially out of one hemisphere and into the other, so hemisphere by hemisphere it weighs as much as a magnetic monopole, and the gravity of a magnetic monopole is known. It is the magnetic Reissner--Nordstr\"om (RN) metric,
\begin{equation}
 ds^2=-h\,dt^2+h^{-1}dr^2+r^2 d\Omega^2 ,
 \qquad
 h=1-\frac{2M}{r}+\frac{P^2}{r^2},
 \label{eq:rn}
\end{equation}
in units $G=c=1$. Glue two copies of it, with magnetic charge $+P$ above the equator and $-P$ below. Both sides carry the same metric, so the junction is smooth; the equatorial plane of this geometry is totally geodesic, so nothing has to be inserted there to hold the two halves apart and the sheet between them carries no surface stress-energy at all. What it must carry is an azimuthal current, because the radial field reverses across it, and once the hole spins a radial current as well, closing the jet's circuit. Those are the two things a disk does to the field it anchors, and this sheet does them and nothing besides: it has no mass, no thickness and no pressure, and it is the source of the entire field (Fig.~\ref{fig:anatomy}).

The flux stands in $h$ alongside the mass, so it is a parameter of the geometry and not a label on a test field. Since the horizon sits where $h$ vanishes, a hole carrying flux is smaller than one of the same mass without it, and enough flux closes the horizon off altogether.

The flux threading each horizon hemisphere is $\Phi=2\pi P$, while the two hemispheres cancel and no magnetic charge is seen from infinity. A horizon exists only for $P\le M$, so a hole can anchor at most
\begin{equation}
 \Phi\le2\pi M .
 \label{eq:bound}
\end{equation}
The bound confines $P/M$ to $[0,1]$. Write $p=P/M$ for that dimensionless flux, whose square is the share of the hole's extremality budget that the field takes up; existing analytic theory is the single point $p\to0$, and this solution covers the whole interval. The construction survives rotation. The same gluing on magnetic Kerr--Newman keeps the equator totally geodesic, so the sheet stays weightless at every spin and the flux stays $\Phi=2\pi P$. There is no Meissner suppression: near-extremal holes expel a field imposed from outside, but this one is sourced by the sheet. The bound sharpens to $\Phi\le2\pi\sqrt{M^2-a^2}$. That member of the family is the electrovac one, the engine's vacuum limit at fixed $M$, $a$ and $P$.

The force-free branch parts from it as soon as the hole turns, and what separates the two is plasma. An accreting hole is never short of it, and force-free is the statement that its stress-energy is negligible beside the field's, so the field is still the only source in the Einstein equations. Rotation adds a frame-dragging rate $w(r)$ to the metric and gives the field lines an angular velocity $\OmF$, and the current the plasma carries gives the field a toroidal component. A degenerate field carries no electric field along its lines, and that forces every flux surface to turn at one rate; for the split monopole the rate is a single constant $\OmF$ for the whole magnetosphere at first order in rotation. The electrovac member fails that test as soon as it spins, where $F_{\mu\nu}\tilde F^{\mu\nu}\propto a$, so the two branches part company at $O(a)$ and meet only at $a=0$. The rotating solution is therefore expanded about the static member of Equation~\ref{eq:rn}, not about Kerr--Newman. The rest is solved rather than assumed: the Einstein, Maxwell and force-free equations are imposed together, to first order in rotation and exactly in the flux. Appendices~\ref{app:static}--\ref{app:working} set that out.

\section{Why a jet cannot be steady}\label{sec:steady}

A magnetosphere is free to choose the poloidal current it carries, and self-gravity takes that freedom away. On a fixed background the drag, the field-line rotation and the toroidal amplitude are only loosely tied, so the current can be set by hand, and that is what lets a test-field magnetosphere be built at all. Here the field equations close on the drag alone,
\begin{equation}
 h\,\big(r^4 w'\big)'=4P^2\big(w-\OmF\big),
 \label{eq:drag}
\end{equation}
and fix the toroidal field at every radius, so the current is not chosen but settled by the geometry.

That freezing is a theorem, and it needs no expansion in rotation. Let the field be degenerate with flux function $\Psi$, constant on each field line, and let the inner boundary be a Killing horizon with generator $\xi$ and angular velocity $\OmH$. Conservation makes the torque carried by a flux tube the same on every sphere, so its value anywhere is its value at the horizon. There the generators neither expand nor shear, because that is what a stationary horizon is. Raychaudhuri's equation then gives $R_{\mu\nu}\xi^\mu\xi^\nu=0$, with no energy condition assumed, and the Einstein equation turns this into $T_{\mu\nu}\xi^\mu\xi^\nu=0$: nothing crosses an equilibrium horizon. For a degenerate field that quantity is known,
\begin{equation}
 4\pi\,T_{\mu\nu}\xi^\mu\xi^\nu\big|_{\mathcal H}
 =\big(\OmF-\OmH\big)^2\big|d\Psi\big|^2 ,
 \label{eq:heatexact}
\end{equation}
non-negative, and zero only where no flux threads or the field corotates with the horizon. Wherever flux does thread the horizon, $\OmF=\OmH$: the field corotates, the current vanishes, and so does the Poynting flux.

In words, a stationary horizon cannot be heated, but a magnetosphere that slips relative to it necessarily heats it, at a rate quadratic in the slip. In the membrane picture the chain is concrete: current crossing the horizon dissipates, the dissipation is heat, the heat is area, and the area of a stationary horizon does not change. The obstruction is therefore a thermodynamic one, about entropy rather than about energy. A test field escapes it because it is allowed to dissipate on a horizon that is not allowed to respond: nothing requires it to satisfy the Einstein equation there, and Equation~\ref{eq:heatexact} is exactly that requirement. Nothing here is assumed at infinity, and at first order in rotation the same conclusion returns as a statement about Equation~\ref{eq:drag}, whose second local solution puts a curvature singularity on the horizon (Fig.~\ref{fig:rigid}, Appendix~\ref{app:first}).

Extraction must therefore be time dependent: a working jet is a black hole in decay, not in equilibrium. What forbids the steady state also fixes the decay. Slow time dependence lifts the obstruction in a boundary layer at the horizon whose transients drain at the surface-gravity rate, so the layer keeps no memory. The freedom gravity removed returns as the hole's own decay, the current constant being the secular drift of the frame dragging (Appendix~\ref{app:working}). Horizon regularity then fixes the current in terms of $\OmH-\OmF$, and the outgoing-wave condition at large radius fixes the same current to the Michel wind value \citep{1973ApJ...180L.133M}. The two agree for one value of $\OmF$ only, and on every field line:
\begin{equation}
 \OmF=\tfrac12\OmH ,
 \qquad
 P_{\rm BZ}=\frac{\kappa_0}{4\pi}\,\Phi^2\OmH^2 ,
 \qquad
 \kappa_0=\frac{1}{6\pi} .
 \label{eq:power}
\end{equation}
The classical impedance match and its coefficient both survive self-gravity unrenormalized at leading adiabatic order; the spin-down shifts the match to $\OmF/\OmH=\tfrac12[1+p^2/6+O(p^4)]$ at the next order, leaving $\kappa_0$ unchanged through $O(p^2)$ (Appendix~\ref{app:ceilingBR}). The rates are outputs rather than inputs. Evaluating the azimuthal Einstein constraint at the horizon returns the torque with no further assumption, $\dot J_H=\tfrac23P^2(\OmF-\OmH)$, which is exactly $-P_{\rm BZ}/\OmF$: the horizon limit of a field equation is the Blandford--Znajek torque, and the rates follow,
\begin{equation}
 \dot M=-P_{\rm BZ},
 \qquad
 \dot J=\dot M/\OmF=-2P_{\rm BZ}/\OmH ,
 \label{eq:rates}
\end{equation}
so the working engine is a geometry of slowly decreasing mass, and a jet is the draining of a rotational reservoir that only the disk's rebuilding of flux sustains.

\section{What a hole can hold, and what it can pay}\label{sec:budget}

\begin{figure}[t]
\centering
\includegraphics[width=0.99\linewidth]{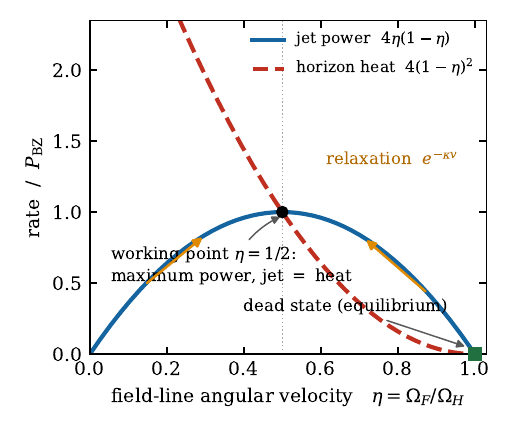}
\caption{The engine as a heat engine. Horizontal axis: the field-line angular velocity in units of the horizon's, $\eta=\OmF/\OmH$. In units of $P_{\rm BZ}$ the jet power is $4\eta(1-\eta)$ and the horizon heating rate $T_H\dot S$ is $4(1-\eta)^2$, so both equal one at the working point. The dead state $\eta=1$ is the unique equilibrium the rigidity theorem allows: zero power, zero dissipation. Horizon regularity and the outgoing-wind condition fix $\eta=1/2$, while the layer drains transients at the surface-gravity rate; there the two curves cross, so the jet's luminosity is matched by the heat the horizon takes.}
\label{fig:cycle}
\end{figure}

Once the engine is a solution, the laws of black-hole mechanics apply to it. The flux of a chargeless, disk-fed hole enters the first law as a charge,
\begin{equation}
 dM=\frac{\kappa}{8\pi}\,dA+\frac{\phi_H}{2\pi}\,d\Phi+\OmH\,dJ ,
 \label{eq:firstlaw}
\end{equation}
with $\kappa$ the surface gravity and $\phi_H=Pr_+/(r_+^2+a^2)$ the horizon magnetic potential. The second law enters through the extracting state. The torque liberates free energy at the rate $\OmH|\dot J|$, and that energy divides in two: half leaves along the jet, half is deposited as horizon area. At leading adiabatic order, the horizon heats at the jet's luminosity, $T_H\dot S=P_{\rm BZ}$. The dead state is the equilibrium and extraction is relaxation toward it. The impedance match that fixes $\OmF$ is a boundary-value statement, not an extremum principle, and the power is stationary there as a consequence (Fig.~\ref{fig:cycle}): a black hole runs its jet at maximum power and fifty percent efficiency.

\begin{figure}[t]
\centering
\includegraphics[width=0.92\linewidth]{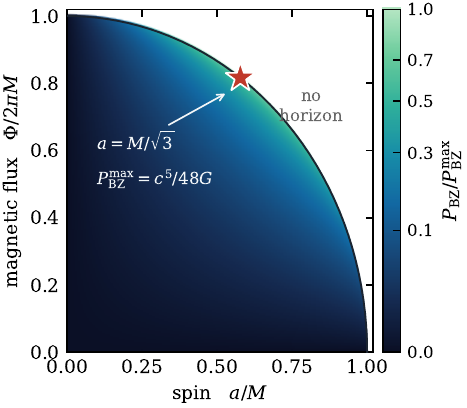}
\caption{Spin and flux share one extremality budget. The horizon exists only inside the quarter disk $\Phi^2+(2\pi a)^2\le(2\pi M)^2$; outside it the two together would overclose the hole. Color: BZ power over the budget, $P_{\rm BZ}\propto\Phi^2\OmH^2$, on a stretched scale (color $\propto$ value$^{0.45}$) so that the bright band near the boundary is legible; the colorbar ticks carry the stretch. The power is maximal on the extremal boundary at $a=M/\sqrt3$ (star), where the budget splits two thirds to flux and one third to spin.}
\label{fig:budget}
\end{figure}

Spin and flux then draw on one account, since both push a horizon toward extremality and here the second is supplied by the disk. Horizon existence is a property of the even sector, so it is the electrovac member, exact at every spin, that fixes where the boundary lies; there the sharpened form of Equation~\ref{eq:bound} is an inequality in spin and flux together,
\begin{equation}
 \Phi^2+(2\pi a)^2\le(2\pi M)^2 ,
 \label{eq:budget}
\end{equation}
so a hole that spins faster can hold less flux. The power depends on both and therefore has a maximum. Maximizing Equation~\ref{eq:power} over the budget is elementary and lands on the extremal boundary at $a^2/M^2=1/3$ (Fig.~\ref{fig:budget}):
\begin{equation}
 P_{\rm BZ}^{\max}=\frac{1}{48}\,\frac{c^5}{G} ,
 \label{eq:ceiling}
\end{equation}
independent of mass, and equal to $7.6\times10^{57}\,{\rm erg\,s^{-1}}$. In circuit language that ceiling is a battery: writing Equation~\ref{eq:power} as $P_{\rm BZ}=(\OmH\Phi/2\pi)^2/6$ identifies $\OmH\Phi/2\pi$ as the electromotive force, which at the optimum is $\sqrt2/4$, or $3.7\times10^{26}\,$V, with a matched current of $4.1\times10^{24}\,$A and $45\,\Omega$ presented by the source and by the load alike. None of these depends on the mass either. The most powerful black hole is extremal yet spins at only $a\simeq0.58M$, because two thirds of its extremality budget goes to flux. The optimum sits where $\kappa=0$, so Equation~\ref{eq:ceiling} is a supremum approached and not attained; it is the supremum of the electrovac budget, and the force-free state's own weight relaxes that budget rather than tightening it (Appendix~\ref{app:extremal}). Evaluating $\kappa_0$ at the optimum stretches the slow-rotation expansion that produced it; solving the force-free field at finite spin on the same geometry multiplies Equation~\ref{eq:ceiling} by $0.8386(2)$ and moves the optimum to $a=0.553M$ (Appendix~\ref{app:ceilingNL}), while the metric back-reaction there is of order unity and of undetermined sign (Appendix~\ref{app:ceilingBR}). At the next order in rotation the power is enhanced rather than suppressed, and on this geometry self-gravity reduces that enhancement without reversing it, the two second-order coefficients being locked in the ratio $8/5$ in the probe limit and drifting from it with flux (Appendix~\ref{app:second}).

The ceiling caps the rate, and because the engine must decay there is also a cap on the total. A calculation on a background with fixed parameters has none, since its energy reservoir does not evolve. If we follow the hole down in spin with $J=aM$, the time drops out of the rates and the trajectory closes, $dM/da=Ma/(2r_+^2+a^2)$. Keeping the leading-order match $\OmF=\tfrac12\OmH$ along the whole spin-down, from an extremal start the probe limit is elementary, $M_f/M_0=e^{1/4}/\sqrt2$: a maximally spinning hole delivers $9.2\%$ of its mass over its entire life while its horizon area grows by exactly $\sqrt e$. Carrying the flux in $r_+$ lowers that to $4.7\%$ at the flux that maximizes the power, since the flux that makes the jet possible takes up part of the same budget (Appendix~\ref{app:extremal}). Because the horizon takes as much heat as the jet takes energy, the work done by the torque, $\int\OmH|dJ|$, is twice the mass delivered, $18.4\%$. The whole $29.3\%$ of rotational energy that Christodoulou's bound makes available is used up: $9.2\%$ leaves as jet and the rest raises the irreducible mass. The total is therefore limited by the reservoir the hole starts with, not by the persistence of the disk that feeds it, and self-gravity lowers both the rate and the reservoir.

\section{The maximal flux limit}\label{sec:endpoint}

Loaded with all the flux it can hold, the hole hands its spin to its own field. At maximal flux the corotating dead state is extremal, so it has a near-horizon limit, and that limit is the Bertotti--Robinson universe, ${\rm AdS}_2\times S^2$ with a uniform field. There the drag equation becomes a conformal-weight problem, $\Delta(\Delta-1)=4$, whose root is irrational,
\begin{equation}
 \Delta=\frac{1+\sqrt{17}}{2}\simeq2.5616 ,
 \label{eq:weight}
\end{equation}
and the irrationality comes from the force-free condition rather than from the field reversal. That single $\OmF$ renders the drag equation homogeneous and produces Euler exponents; in vacuum the same sector has integer ones (Appendix~\ref{app:extremal}). Toward extremality the weight governs the horizon's own moment of inertia, the angular momentum it holds per unit of its rotation rate. That moment vanishes as $\sigma^{\Delta-1}$, with $\sigma$ the remaining distance to extremality, while the magnetosphere's does not; a Kerr--Newman horizon, by contrast, keeps at least a third of the total angular momentum all the way to extremality. The two limits are taken in order, first the linear response and then $\sigma\to0$, and what survives is the ratio. In this limit the dead state becomes magnetosphere dominated: the rotation resides in the field the hole drags rather than in the horizon itself. The result concerns the corotating, zero-power branch as the flux approaches its bound.

\section{Scope of the results}\label{sec:scope}

It is worth setting out how much of this is proved and how much is computed. Three results are exact: the static solution with its sheet, at every flux, and the flux bound of Equation~\ref{eq:bound} and the budget of Equation~\ref{eq:budget}, at every spin and flux, as statements about the electrovac member of the family. The rigidity theorem is exact in a stronger sense still, since it needs no expansion in rotation and no condition at infinity; Appendix~\ref{app:first} lists the hypotheses it does need. The first-order solution is exact in structure rather than in closed form: the reduction is analytic and the toroidal field is locked to the drag algebraically, but the drag profile itself solves a two-point boundary value problem whose homogeneous equation admits no Liouvillian solution at any $0<p<1$ (Appendix~\ref{app:first}). The impedance match, the rates and the heat are controlled at first order in rotation and to leading adiabatic order, for the split-monopole class with a smooth horizon and no bulk disk stress. Their relative corrections are of order $(M\OmH)^2$ and $\Gamma_J/\kappa$, the fractional spin-down rate in units of the surface gravity, which is $p^2/3$ at small flux, $1.4\%$ at $p=0.2$, and reaches unity at $p\simeq0.85$. What remains conjectural is the budget beyond the split class, and whether the maximal flux limit can be reached in finite time. The principal results have been checked against independent benchmarks or by internal consistency checks, and Appendix~\ref{app:verify} lists the checks and what they returned. Two companion papers carry the work further. One pushes the rotation expansion to exact coefficients through at least eighth order. The other applies the solution to gamma-ray bursts, where the timescales and the energies it gives are consistent with the observed ones. The flux here comes from the disk, idealized as the sheet. Black-hole vortices that trap magnetic flux are another possible configuration \citep{2022PhRvL.129f1302D}; we treat the disk first because it is also the most natural source of the force-free plasma.

\section{Conclusion and outlook}\label{sec:conclusions}

Blandford and Znajek gave the mechanism in 1977, and for fifty years it has been sharpened, simulated and compared with images of real black holes, always on a spacetime that was fixed in advance and never responded to the field threading it. Here that field is allowed to gravitate, and quantities that used to be put in by hand come out instead. The current the magnetosphere carries is no longer free to choose; the geometry fixes it. A jet cannot be steady, because a hole in equilibrium cannot absorb the heat that a working magnetosphere delivers to it, so extraction is a slow decay whose rate the equations set. And the magnetic flux the jet needs pushes the hole toward the same limit that spin does, so the two must share a single budget. That shared budget is what caps the power, at $c^5/48G$, at leading order in rotation, however heavy the hole, and what holds the energy delivered over an entire lifetime, in the leading-order evolution, below a tenth of the hole's mass. Through all of it the original answer survives intact, the field lines still turning at half the rate of the horizon, up to a retardation correction of relative size $p^2/6$ from the spin-down, now derived rather than assumed.

The construction is no more than two textbook geometries glued along a current sheet, and it works because the split monopole is the one field whose weight can be carried exactly. That is also its limit. The results hold for this idealized configuration, the simplest theoretical approximation to an accretion disk and the jet. Whether the budget, the ceiling and the decay survive for a more realistic disk-anchored field is the next question, and it is now one that can be put, since there is something exact to perturb around. Whether a hole can reach the maximal flux limit in finite time is open in the same way. The answers above are the first of a class of questions that could not be asked before, and probably the smaller part of what the solution will eventually yield.

\begin{acknowledgments}
The author thanks Yi-Zhong Fan and Rongrong Xue for the invitation to contribute to the inaugural issue of \textit{Astronomy Communications}, and hopes it will grow into one of the influential and enduring journals of the field.
Anthropic's Claude was used to optimize the language of the manuscript and to assist with the coding, and the author has verified both.
\end{acknowledgments}

\appendix
\renewcommand{\theHequation}{\thesection.\arabic{equation}}

\section{The static engine and its disk}\label{app:static}

Conventions: $G=c=1$, signature $(-,+,+,+)$, $\nabla_\nu F^{\mu\nu}=4\pi J^\mu$, azimuthal period $2\pi$.

\subsection{The solution}
Take the magnetic RN metric of Equation~\ref{eq:rn} with the split-monopole field
\begin{equation}
 F^{(0)}=\varsigma(\theta)\,P\sin\theta\,d\theta\wedge d\varphi ,
 \qquad
 \varsigma\equiv\mathrm{sign}(\cos\theta) .
 \label{eq:F0}
\end{equation}
In each open hemisphere this is the magnetically charged RN electrovac pair and the Einstein--Maxwell equations hold exactly. The stress tensor is quadratic in $F$, so $\varsigma$ drops out of $T_{\mu\nu}$ and the two hemispheres carry the same metric. The equatorial plane of RN is totally geodesic, so the Israel junction across $\theta=\pi/2$ has zero jump in extrinsic curvature: the sheet carries no surface stress-energy. The tangential magnetic field jumps by $[B_{\hat r}]=2P/r^2$, which requires the azimuthal surface current
\begin{equation}
 4\pi K_{\hat\varphi}=\frac{2P}{r^2} .
 \label{eq:Kphi}
\end{equation}
The flux through a large sphere vanishes, so no magnetic charge is seen from infinity, while hemisphere by hemisphere the flux $\Phi=2\pi P$ threads the horizon. The sheet carries the azimuthal current that sources the poloidal field, and, once the hole spins, the radial return current that closes the BZ circuit,
\begin{equation}
 4\pi K_{\hat r}=\frac{\OmH P}{r\sqrt{h(r)}},
 \qquad
 4\pi K^{r}=\frac{\OmH P}{r},
 \label{eq:Kr}
\end{equation}
this last current being the force-free branch's, fixed by the Znajek relation of Appendix~\ref{app:first}. Its $h^{-1/2}$ growth is a static-frame artifact: the coordinate component is finite, and the ingoing component $4\pi K^{v}=2P(\OmH-w)/rh$ is regular, the Znajek relation making its numerator vanish exactly where $h$ does. For the electrovac glue of the next paragraph the invariant norm is clean at every spin, $|K|^{2}=P^{2}/4\pi^{2}r^{4}$. A conducting equatorial disk is in any case a prerequisite of the split-monopole topology; without one the sheet reconnects \citep{2004MNRAS.350..427K}. Here that surface is promoted to the gravitating source.

\subsection{The probe wind survives}
For arbitrary $\OmF(\Psi)$ and $s=\pm1$, the field
\begin{equation}
 f=\Psi_{\rm m}\sin\theta\,d\theta\wedge(d\varphi-\OmF dt)
 +s\,\frac{\OmF\Psi_{\rm m}\sin\theta}{h}\,dr\wedge d\theta
 \label{eq:michel}
\end{equation}
solves the full nonlinear force-free system on Equation~\ref{eq:rn} exactly: the RN extension of Michel's rotating monopole \citep{1973ApJ...180L.133M}. The coupled analysis promotes Equation~\ref{eq:michel} with $\Psi_{\rm m}=P$ from probe to source.

\subsection{All spins}
Gluing opposite-charge magnetic Kerr--Newman across the equator keeps $\partial_\theta g_{ab}=0$ there, so the equator stays totally geodesic and the sheet stays massless at every spin. The sheet currents are $4\pi K^t=2Pa/r^3$ and $4\pi K^\varphi=2P/r^3$, and the hemispheric flux is $\Phi=2\pi P$ with no Meissner suppression, bounded by horizon existence as $\Phi\le2\pi\sqrt{M^2-a^2}$, which is the budget of Equation~\ref{eq:budget}. The price is that the bulk field is not force-free:
\begin{equation}
 F_{\mu\nu}\tilde F^{\mu\nu}
 =-\,\frac{8P^2a\,r\cos\theta\,(r^2-a^2\cos^2\theta)}
          {(r^2+a^2\cos^2\theta)^4} ,
\end{equation}
which vanishes on the equator and at $a=0$, and at fixed radius is largest on the axis whenever $r>3.08\,a$, hence everywhere in the slow-rotation regime. This member of the family is the engine's vacuum limit at fixed $M$, $a$ and $P$, the same hole with the same flux and the same spin and no plasma in its magnetosphere: it is electrovac, so it carries no bulk current, and with none it carries no toroidal field and no Poynting flux. The force-free branch parts from it at first order in rotation, where degeneracy replaces the Kerr--Newman profile $a/r^2$ by a constant $\OmF$, and the two meet only at $a=0$, where the purely magnetic split monopole is degenerate and carries no current in the bulk. The force-free rearrangement of this field is exactly what the slow-rotation expansion below computes.

\section{First order in rotation: reduction and rigidity}\label{app:first}

\subsection{Ansatz and closed-form reduction}
Let $\epsilon$ count rotation,
\begin{align}
 ds^2&=ds^2_{\rm RN}-2\epsilon\,w\,r^2\sin^2\theta\,dt\,d\varphi
 +2\epsilon\,\beta\,dr\,d\varphi ,\\
 F&=F^{(0)}+\epsilon\varsigma\big[\OmF P\sin\theta\,dt\wedge d\theta
 +u\,dr\wedge d\theta\big] .
\end{align}
The single nontrivial force-free component closes on the gauge-invariant toroidal combination $\hat u\equiv u+P\beta/(r^2\sin\theta)$, which is what survives the residual freedom $\varphi\to\varphi+\epsilon\zeta(r,\theta)$: it gives $r^3h\,\hat u_{,r}+2(Mr-P^2)\hat u=0$, i.e.\ $\hat u=\mathcal C(\theta)/h(r)$, with poloidal current $I=\tfrac12 \mathcal C(\theta)\sin\theta$. The drag does not enter. In horizon-penetrating coordinates built from the principal null congruence of the perturbed metric, which requires the azimuthal shift with $\partial_r\zeta=w/h+\beta/(r^2\sin^2\theta)$ rather than the naive $\partial_r\zeta=w/h$, the coefficient of $dr\wedge d\theta$ is $\epsilon[\mathcal C-(\OmF-w)P\sin\theta]/h$, so future-horizon regularity forces the Znajek relation
\begin{equation}
 \mathcal C(\theta)=(\OmF-\OmH)P\sin\theta ,
 \qquad \OmH\equiv w(r_+) .
 \label{eq:znajek}
\end{equation}
The only nontrivial Einstein components are $(t\varphi)$, $(r\varphi)$, $(\theta\varphi)$, and they say something sharper than a reduction. Writing $E_{\mu\nu}$ for the residual of the Einstein equation, the $(r\varphi)$ row is $E_{r\varphi}=2P\sin\theta\,\hat u/r^2$: it states that the gauge-invariant toroidal field, and with it the current, \emph{vanishes} in the $\ell=1$ sector, which with Equation~\ref{eq:znajek} already forces $\OmF=\OmH$. The $(\theta\varphi)$ row holds identically for $\beta\propto\sin^2\theta$, and the $(t\varphi)$ row closes on $w$ alone, giving Equation~\ref{eq:drag}. Two checks anchor the system: at $P=0$ it returns the vacuum drag $w=2J/r^3$, and the slow-rotation vacuum Kerr--Newman field, whose profile is $a/r^2$ in place of $\OmF$, satisfies it identically. A current-carrying configuration therefore takes a more general form than $\beta=Q(r)\sin^2\theta$; solving the two constraints without that restriction gives $h\beta=\mathcal G(r)\sin^2\theta+\mathcal M(\theta)$ with $(\partial^2_\theta-\cot\theta\,\partial_\theta+2)\mathcal M=4P\sin\theta\,\mathcal C$, whose source is resonant against $\sin^2\theta$. The ingoing formulation of Appendix~\ref{app:working} avoids this bookkeeping entirely, and it is the one we rely on: there the lock of Equation~\ref{eq:lock} is exactly the statement $I\equiv0$, and $I=\tfrac12\mathcal C_0\sin^2\theta$ once the current constant is restored.

\subsection{Completeness}
The reduction is forced. Take the most general stationary axisymmetric odd-parity sector at $O(\epsilon)$, with metric functions $H_{t\varphi},H_{r\varphi},H_{\theta\varphi}$ and field components $F_{tr},F_{t\theta},F_{r\theta}$ all free functions of $(r,\theta)$. Degeneracy $F\wedge F=0$ forces $F_{tr}=0$, and closedness then gives the isorotation form $F_{t\theta}=\OmF(\theta)P\sin\theta$; force-freeness closes on the gauge-invariant toroidal combination $\hat u\equiv u+P H_{r\varphi}/(r^2\sin\theta)$ and returns $\hat u=\mathcal C(\theta)/h$; the $(t\varphi)$ equation carries the angular operator $\partial_\theta^2-\cot\theta\,\partial_\theta$, whose regular eigenfunctions organize the system into multipoles. Per multipole the Frobenius resonance condition reads $[(\ell(\ell+1)-2)r_+^2+4P^2]A_\ell(r_+) +4P^2r_+^2(\OmF\sin^2\theta)_\ell=0$, with $A_\ell$ the $\ell$th multipole of $H_{t\varphi}$. At $\ell=1$ the bracket is $4P^2$ and the condition is $w(r_+)=\OmF$, which is the theorem. For $\ell\ge2$ it gives $w_\ell(r_+)\neq\Omega_{F,\ell}$, so horizon smoothness alone leaves rigid rotation open for a general $\OmF(\theta)$; the zeroth law closes it. Requiring the horizon to be a Killing horizon makes $w(r_+,\cdot)$ constant, which feeds back to kill every $\ell\ge2$ component of $\OmF\sin^2\theta$ and leaves $\OmF={\rm const}=\OmH$, hence $\mathcal C=0$, without invoking the far-field conditions at all. Alternatively, if $\OmF$ is taken constant at the outset, the horizon relation becomes $LY=(4P^2/r_+^2)Y$ for $Y=w(r_+,\cdot)-\OmF$ with $L=\partial^2_\theta+3\cot\theta\,\partial_\theta$, and $L$ is negative semidefinite in the weight $\sin^3\theta$, so $Y\equiv0$.

\subsection{The rigidity theorem}
Equation~\ref{eq:drag} has a regular singular point at $r_+$ with indicial roots $\{0,1\}$ and nonvanishing resonance residue $4r_-/[r_+(r_--r_+)]$, with $r_-$ the inner horizon, so its two local solutions behave as $w-\OmF\sim{\rm const}\times h$ and $w-\OmF\sim{\rm const}+{\rm const}\,(r-r_+)\ln(r-r_+)$. The second has divergent $w'$. Since $w=-g(\partial_t,\partial_\varphi)/g(\partial_\varphi,\partial_\varphi)$ is a scalar built from the Killing vectors, the logarithmic branch cannot be removed by any chart, ingoing Eddington--Finkelstein included. The problem runs deeper than a divergent derivative: in the regular ingoing chart twelve Riemann components carry $w''$, among them $R^\varphi{}_{rvr}=\tfrac12w''$, and $w''\sim(r-r_+)^{-1}$ on that branch, so the horizon carries a curvature singularity. The statement has to be made about components: by $t$--$\varphi$ parity the $O(\epsilon)$ Kretschmann scalar vanishes identically, so no scalar invariant sees the branch at this order. Hence $\OmH=w(r_+)=\OmF$, and $\mathcal C=0$: current, toroidal field, horizon torque and Poynting flux all vanish.

\begin{figure}[t]
\centering
\includegraphics[width=0.9\linewidth]{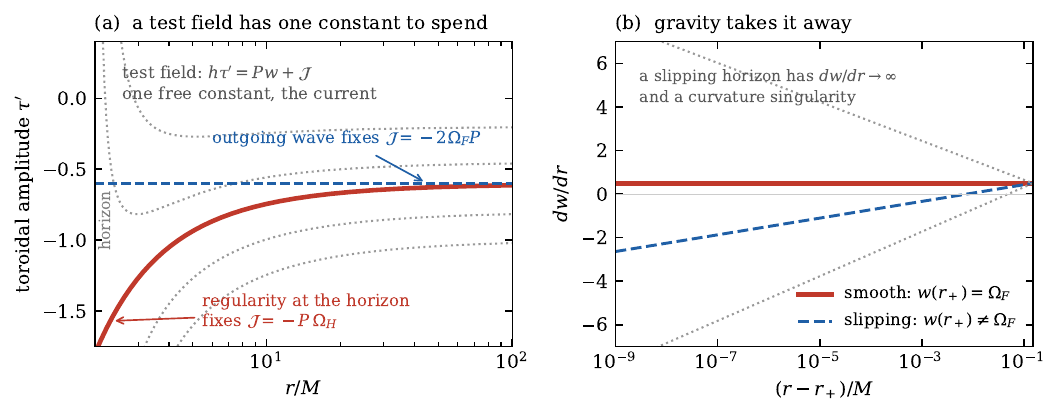}
\caption{Why a jet cannot be steady. \textit{(a)} On a fixed metric the toroidal amplitude obeys $h\tau'=Pw+\mathcal J$, a one-parameter family; $\mathcal J$ is the current the magnetosphere runs on. Only one member (red) is regular at the horizon, which fixes $\mathcal J=-P\OmH$, and its value at infinity must match the outgoing Michel wind, $\mathcal J=-2\OmF P$. The two conditions together give $\OmF=\OmH/2$: this is the classical impedance match, and it is possible only because $\mathcal J$ was free to be chosen. \textit{(b)} Once the field gravitates the Einstein equations leave no such constant. The drag obeys Equation~\ref{eq:drag}, whose smooth branch has $w(r_+)=\OmF$ and finite slope (red), while every solution with a slipping horizon has $dw/dr\to\infty$ logarithmically (blue), and with it a curvature singularity.}
\label{fig:rigid}
\end{figure}

\subsection{The same theorem without the expansion}
Let $\xi_{(t)}=\partial_t$ and $\xi_{(\varphi)}=\partial_\varphi$ be the Killing fields, let $F$ be degenerate with flux function $\Psi$, and let the inner boundary of the domain of outer communication be a Killing horizon $\mathcal H$ with generator $\xi=\xi_{(t)}+\OmH\xi_{(\varphi)}$. Since $\xi_{(\varphi)}$ is Killing and $\nabla_\mu T^{\mu\nu}=0$, the current $j^\mu=-T^\mu{}_\nu{\xi_{(\varphi)}}^\nu$ is divergence free, and degeneracy makes it tangent to the flux surfaces, $j^\mu\partial_\mu\Psi=F^{\mu\nu}\partial_\mu\Psi\partial_\nu\Psi/4\pi=0$; the torque of a flux tube is therefore the same on every sphere. This is a statement about flux and not about contained charge, and the distinction is what makes it usable, since the dead state's quasi-local angular momentum grows linearly in $r$ and any argument phrased as ``the angular momentum inside a sphere is finite and constant'' fails at the first line. On $\mathcal H$ the induced metric is Lie dragged along $\xi$, so expansion and shear vanish identically, Raychaudhuri gives $R_{\mu\nu}\xi^\mu\xi^\nu=0$ with no energy condition, and Einstein turns this into $T_{\mu\nu}\xi^\mu\xi^\nu=0$. For a degenerate field $\iota_\xi F=(\OmF-\OmH)d\Psi$, so
\begin{equation}
 4\pi\,T_{\mu\nu}\xi^\mu\xi^\nu\big|_{\mathcal H}
 =\big(\OmF-\OmH\big)^2\big|d\Psi\big|^2_{\mathcal H} ,
 \label{eq:horizon-heat-exact}
\end{equation}
forcing $\OmF=\OmH$ wherever flux threads the horizon, hence $I=0$ and zero torque at every radius. The hypotheses of the theorem are: two commuting Killing fields under which $F$ is invariant, a regular axis, an inner boundary that is a Killing horizon with $g$ and $F$ smooth on it in a horizon-penetrating chart, a degenerate field, poloidal flux actually threading the horizon, and no current sheet transmitting a radial torque. The conclusion holds on the field lines that thread the horizon, so the theorem bounds what can be taken from the hole and says nothing about a disk-anchored outflow.

\citet{1977MNRAS.179..433B} already observed that the irreducible mass is constant only when $\OmF=\OmH$, and \citet{1977MNRAS.179..457Z} that a hole with $\OmF\neq\OmH$ dissipates; the step from $\OmF=\OmH$ to $I=0$ and zero flux is a theorem in \citet{1982MNRAS.198..345M} and \citet{2014MNRAS.445.2500G}, though only for closed loops, both ends on the horizon, where it forbids ingrown hair rather than jets. What is added here is the inversion: on a fixed background those statements say how fast a hole is heated, and once the field gravitates a Killing horizon cannot be heated at all, so the same expression becomes an equation for $\OmF$. Equation~\ref{eq:horizon-heat-exact} is a statement about $T_{\mu\nu}\xi^\mu\xi^\nu$ and not about a horizon torque, so it is untouched by the objection of \citet{2014PhRvD..89b4041L} to membrane-paradigm torque language. Carter's circularity theorem \citep{1969JMP....10...70C} is the other statement a reader will reach for; its hypothesis is the absence of a poloidal flux of energy and angular momentum, and it fails for a magnetosphere carrying poloidal current \citep{2012LRR....15....7C}, which is why a working engine is necessarily non-circular and why the corotating dead state, circular and currentless, is the member the classical theorems do reach.

\subsection{The dead state}
Setting $y=w-\OmF$, Equation~\ref{eq:drag} becomes $h(r^4y')'=4P^2y$ with $y(r_+)=0$ and $y\to-\OmF$ at infinity. At leading order in $p$ this gives $w=8M^3\OmF/r^3$ and $J=4M^3\OmH$; the $O(p^2)$ correction is closed form and gives the moment of inertia $J_H/(4M^3\OmH)=1-\tfrac{10}{9}p^2+O(p^4)$, which a direct solve of the boundary value problem reproduces to ten significant digits ($-1.111111111$ against $-10/9$, with $O(p^4)$ coefficient $+0.10792$). The drag acquires the slow tail $w\to2\OmF P^2/r^2$, which is why the theorem is stated with no condition at infinity: $g_{t\varphi}$ approaches a constant times $\sin^2\theta$ instead of falling off, and the quasi-local $J(r)=-\tfrac16r^4w'$ grows linearly, the near-zone signature of the light cylinder at $r\sim1/\OmF$, beyond which the slow-rotation expansion does not reach. The homogeneous equation is provably transcendental: the monodromy contains an infinite-order unipotent element, so no solution is algebraic; the horizon points are not apparent singularities; a direct search empties the finite-dimensional space of rational candidates; and the Kovacic algorithm, which decides whether a second-order linear equation admits Liouvillian (closed-form) solutions, excludes them with the flux parameter kept symbolic. Every obstruction closes on the same factor $\sigma^2-1$, with $\sigma=\sqrt{1-p^2}$ the gap to extremality of the static hole, so $P=0$ is the one value at which a closed form exists, and there the rational search returns the vacuum drag; at the extremal endpoint the irrational indicial exponents $(1\pm\sqrt{17})/2$ empty the first Kovacic case at once.

\section{The working engine}\label{app:working}

\subsection{Ingoing form and the lock}
In ingoing coordinates $(v,r,\theta,\varphi)$ the static engine is $ds^2=-h\,dv^2+2dv\,dr+r^2d\Omega^2$. At $O(\epsilon)$, after gauging away $g_{r\varphi}$ and $A_r$, the odd sector is $g_{v\varphi}=-w(r)r^2\sin^2\theta$ with $F_{v\theta}$ (coefficient $\OmF$), $F_{r\theta}=\tau'(r)\sin\theta$ from the toroidal potential $A_\theta=\tau(r)\sin\theta$, and the monopole $F_{\theta\varphi}=P\sin\theta$. The complete first-order system is
\begin{align}
 (r^4w')'&=4P\tau', &
 h(r^4w')'&=4P^2(w-\OmF), \nonumber\\
 (h\tau')'&=Pw' , &&
 \label{eq:ingoing}
\end{align}
angular-momentum transport, drag, and induction. The system is overdetermined and its consistency condition is algebraic,
\begin{equation}
 h\,\tau'=P\big(w(r)-\OmF\big) ,
 \label{eq:lock}
\end{equation}
pointwise, with no integration constant: Einstein's equations tie the toroidal profile rigidly to the frame drag. In the probe limit only induction survives, whose first integral $h\tau'=Pw+\mathcal J$ carries the free constant that an astrophysical magnetosphere uses to run. The lock is what freezes when the field must also gravitate.

\subsection{The layer drains}
Let the toroidal sector depend on time, $F_{r\theta}=\tau'(v,r)\sin\theta$. Eliminating the electric amplitude $E$, defined by $F_{v\theta}=E\sin\theta$, between the Bianchi identity $\partial_r E=\partial_v\tau'$ and the azimuthal force-free row, the toroidal field obeys the exact transport equation
\begin{equation}
 2\,\partial_v\tau'+\partial_r(h\tau')=P\,\partial_r w ,
 \label{eq:transport}
\end{equation}
whose stationary solutions are the drag-sourced particular integral plus the one-parameter family $h\tau'={\rm const}$, i.e.\ the lock of Equation~\ref{eq:lock} with the current constant of Equation~\ref{eq:c0drag}. Horizon smoothness selects one member. The characteristics satisfy $dr/dv=h/2\ge0$: they move outward everywhere outside the horizon and are tangent to it at $r_+$, so no characteristic leaves the horizon. Writing $\tau'=\tau'_{\rm stat}+\delta\tau'$, the deviation obeys the homogeneous equation with its source slaved through the drag, and along the horizon generator $d\,\delta\tau'/dv=-\tfrac12h'(r_+)\delta\tau'$: deviations decay as $e^{-\kappa v}$ with $\kappa=\tfrac12h'(r_+)$, the surface gravity, while the outward part is advected into the wind. Numerically, at $M=1$, $P=0.6$ the amplitude ratio after $v=40$ is $5.137\times10^{-5}$, against $e^{-\kappa v}=5.137\times10^{-5}$.

The separation of timescales varies across the family. Relaxation takes $\kappa^{-1}\sim M$, which is faster than the mass drain by $O(\epsilon^{-2})$, while the fractional spin-down rate of the horizon, $\Gamma_J\equiv|\dot J_H/J_H|=P^2/(12M^3I_H)$, with $I_H=J_H/(4M^3\OmH)$ the horizon moment of Appendix~\ref{app:extremal}, is independent of $\epsilon$, so the relevant ratio is $\Gamma_J/\kappa=p^2(1+\sigma)^2/(12\sigma I_H)$, which is $p^2/3$ at small flux. Adiabaticity is therefore controlled by the flux, and it degrades both as $\kappa\to0$ and as the horizon moment falls: $\Gamma_J/\kappa$ is $0.033$ at $p=0.3$, $0.35$ at $p=0.7$ and $1.79$ at $p=0.9$, and crosses unity at $p\simeq0.85$; the weak-flux value $I_H=1$ would underestimate it by a factor $5.6$ at $p=0.9$. The same ratio bounds the departure from isorotation, since $\OmF$ is constant along field lines only up to terms of order $\partial_v$: the working state is isorotating to relative order $\Gamma_J/\kappa$, which is $1.4\%$ at $p=0.2$. The quasi-stationary treatment is controlled when $\Gamma_J/\kappa\ll1$; this leading-order estimate, made with the static linear-response values of $\kappa$ and $I_H$, reaches unity near $p\simeq0.85$, indicating loss of timescale separation there, and the extremal statements of Appendix~\ref{app:extremal} concern the stationary dead state, which needs no such separation.

Including gravity changes nothing about what propagates. The characteristic polynomial of the time-dependent odd sector factorizes as $k_r^2\omega^2(2\omega+hk_r)$ in the characteristic covector $(\omega,k_r)$ along $v$ and $r$, so the speeds are $dr/dv\in\{0,0,h/2\}$: no gravitational mode propagates, $w$ is recovered from $\tau'$ by two radial quadratures at each $v$, and the $(v\varphi)$ row is a constraint whose radial derivative reproduces Equation~\ref{eq:transport} modulo the $(r\varphi)$ row.

\subsection{Michel's rotator and the factor of two}
In ingoing coordinates the exact flat-space rotator has two force-free branches for the toroidal amplitude, $F_{r\theta}\in\{0,-2\OmF P\sin\theta\}$, with radial energy flux $-T^r{}_v=\mp\OmF^2P^2\sin^2\theta/4\pi r^2$. The null amplitude is pure ingoing radiation in these coordinates; the outgoing wind is the second branch, and the factor of two arises because the ingoing component mixes $E_\theta+B_{\hat\varphi}$. The matching chain is then three lines: $h\tau'=Pw+\mathcal J$ from induction; smoothness at the future horizon gives $\mathcal J=-P\OmH$; the far limit $h\tau'\to \mathcal J$ must equal $-2\OmF P$; hence $\OmF=\OmH/2$ and, with $I=\tfrac12\mathcal C(\theta)\sin\theta$ and $dP_{\rm BZ}/d\theta=-\OmF I\Psi_{,\theta}$, the power of Equation~\ref{eq:power}. The classical coefficient survives because the static sector is spherically symmetric, so the $\sin^3\theta$ horizon weighting of the split monopole is untouched.

\subsection{Secular rates}
Degeneracy makes the flux identity $T^r{}_v=-\OmF T^r{}_\varphi$ exact for any member of the wind family: energy flux is $\OmF$ times angular-momentum flux, line by line. On the drifting engine $h=1-2M(v)/r+P^2/r^2$ the only new nonvanishing Einstein component is the constraint $G^r{}_v=2\dot M/r^2$. This is isotropic while $T^r{}_v\propto\sin^2\theta$, so the balance $G^r{}_v=8\pi T^r{}_v$ fixes the drift through its $\ell=0$ projection, with $\langle\sin^2\theta\rangle=2/3$: $\dot M=-\tfrac23\OmF^2P^2=-P_{\rm BZ}$, in agreement with the global integral $\oint(-T^r{}_v)\,dA$. The $\ell=2$ remainder is carried by the anisotropic sector of Appendix~\ref{app:second}. The flux identity then delivers $\dot J=\dot M/\OmF$. Both rates in Equation~\ref{eq:rates} are integrability conditions of the $(r,v)$ and $(r,\varphi)$ constraints.

\subsection{The secular first-order state, and where the current constant comes from}
Restoring slow time dependence in the odd sector, the $(v\varphi)$ Einstein row reads
\begin{equation}
 h(r^4w')'+4P(E-Pw)+r^4\partial_v w'=0 ,
 \label{eq:vphi}
\end{equation}
with $E=\OmF P$ the electric amplitude. The last term is what a strictly stationary treatment drops, and it must be kept: the fractional spin-down rate is independent of $\epsilon$, so $\partial_v w$ is of the same order as the other terms. Writing
\begin{equation}
 \mathcal C_0(r)\equiv-\frac{r^4\partial_v w'}{4P}
\end{equation}
puts Equation~\ref{eq:vphi} in the $\mathcal C_0$-sourced form
\begin{align}
 h(r^4w')'&=4P^2(w-\OmF)+4P\mathcal C_0 , \nonumber\\
 h\tau'&=P(w-\OmF)+\mathcal C_0 .
 \label{eq:c0drag}
\end{align}
The current constant is therefore the secular drift of the frame dragging, related to the first-integral constant $\mathcal J$ of $h\tau'=Pw+\mathcal J$ in Appendix~\ref{app:first} by $\mathcal C_0=\mathcal J+P\OmF$. A uniform drift of the toroidal potential $A_\theta$ is pure gauge, equivalent to a shift of $\OmF$, and cannot play this role.

Evaluating Equation~\ref{eq:vphi} at the horizon, where $h$ vanishes, then gives the torque with no further input:
\begin{align}
 4P^2(\OmF-\OmH)&=-r_+^4\partial_v w'(r_+)=6\dot J_H \nonumber\\
 \Longrightarrow\quad
 \dot J_H&=\tfrac23P^2(\OmF-\OmH),
 \label{eq:torque}
\end{align}
which is exactly $-P_{\rm BZ}/\OmF$ with $P_{\rm BZ}=\tfrac23\OmF(\OmH-\OmF)P^2$. The horizon limit of the $(v\varphi)$ constraint \emph{is} the Blandford--Znajek torque, and it fixes $\mathcal C_0(r_+)=-P(\OmH-\OmF)$, the Znajek value. At that value the right side of the first line of Equation~\ref{eq:c0drag} cancels at $r_+$, the interior problem collapses to $h(r^4w')'=4P^2(w-\OmH)$ with $h\tau'=P(w-\OmH)$, and $w(r_+)=\OmH$ holds on a smooth branch. Note what this does to $\OmF$: it drops out of the interior problem entirely and is fixed only by the far-field branch, which is why the matching of the previous subsection carries the whole determination. The logarithmic branch is an artifact of the strictly stationary equation, and rigidity is unaffected, since strict stationarity means $\mathcal C_0=0$. The constancy of $\mathcal C_0$ is a near-zone statement: at large radius $r^4w_1'\sim-4PE_1r$, with $w_1$ and $E_1$ the drift rates $\partial_v w$ and $\partial_v E$.

\section{Second order and the two-timing composite}\label{app:second}

\subsection{Corotating state}
With the Hartle-type even ansatz the full $O(a^2)$ system closes into twenty radial equations. Three of the results at this order are structural. The electric potential correction decouples identically: degeneracy forces $\partial_r a_{t2}=0$, so the only electric freedom is a uniform shift of $\OmF$ and the impedance is not renormalized at second order. The $(r\theta)$ rows are algebraic in the flux amplitudes and force their symmetric combination to be constant, the $O(\epsilon^2)$ flux shift, which enters $\delta M$ through the $(\phi_H/2\pi)d\Phi$ slot of the first law and vanishes at fixed flux. The $(vv)$ row contains no lapse perturbation, so the Hartle mass-function structure survives and $\delta M$ is a first-order radial integral. At fixed flux that integral is elementary. Writing $y=w-\OmF$ and using the drag equation, the $\ell=0$ row is
\begin{equation}
 \delta m'=\tfrac1{12}r^4y'^2+\frac{P^2}{3h}\,y^2
 =\frac{d}{dr}\Big[\tfrac1{12}\,y\,r^4y'\Big] ,
 \label{eq:deltam}
\end{equation}
so the second-order mass function is known in closed form,
\begin{equation}
 \delta m(r)=\tfrac12\big[\OmH-w(r)\big]\,J(r),
 \qquad J(r)=-\tfrac16r^4w' .
 \label{eq:flywheel}
\end{equation}
This is the first law's rotational term as a pointwise identity. It vanishes at the horizon, where $w=\OmH$, and at $M\ll r\ll1/\OmF$, which is outside the hole and inside the light cylinder, it becomes $\delta M=\tfrac12\OmH J_\infty$: the engine stores rotational energy exactly as a flywheel of moment $J/\OmH$, at every flux, with no fitting. The partner of $\delta M$ is the asymptotic quasi-local angular momentum, and that quantity needs a definition, because $J(r)$ does not converge: the slow tail of the drag makes it grow, and the near-zone expansion of the dead state reads
\begin{equation}
 J(r)=\tfrac23\OmF P^2 r+\tfrac43M\OmF P^2\ln(r/M)+J_\infty
 +O(\ln r/r) ,
 \label{eq:Jinf}
\end{equation}
the linear and logarithmic coefficients following from $1/h$ in the drag equation. We define $J_\infty$ by subtracting them, with the logarithm referred to the scale $M$; the finite part of $\delta M$ below inherits the same subtraction; what is left is constant to six digits between $r=10^5M$ and $10^7M$. So defined, $J_\infty$ differs from the horizon Komar charge at second order in the flux, $J_\infty/J_H=1-0.2310p^2+O(p^4)$, so a fit that identifies the two is already off by $8.5\%$ at $p=0.5$. The tails are consistent with Equation~\ref{eq:flywheel} term by term: the $\ell=0$ radial metric perturbation $k_{20}=\delta g_{rr}h$ tends to $\tfrac23\OmF^2P^2$ with $\ln r/r$ coefficient $\tfrac43M\OmF^2P^2$, while $J(r)$ grows with linear coefficient $\tfrac23\OmF P^2$, the light-cylinder behavior already seen in Appendix~\ref{app:first}.

\subsection{Extracting state}
The even sector at $O(\epsilon^2)$ in the ingoing gauge $g_{rr}=g_{v\theta}=g_{r\theta}=0$ (functions $A,B,U$ and the tensor mode $V$ splitting $\theta\theta$ from $\varphi\varphi$, plus the flux perturbation, each with stationary and drift parts) gives eleven component equations that close at linear drift order, with the force-free consistency rows vanishing identically on the induction integral. Dropping $V$ is an over-restriction that a Kerr-consistency check exposes. The drift is exactly spherical Vaidya: the linear-in-$v$ rows vanish identically on the mass-drift mode, anisotropic drift modes are excluded because their far field decays one power too fast, and the $(r,v)$ constraint fixes
\begin{equation}
 \dot M=\tfrac23\OmF P\big(\OmF P+\tau'_\infty\big) ,
 \label{eq:mdotgen}
\end{equation}
which vanishes for the corotating pair $\tau'_\infty=-\OmF P$ and returns $-P_{\rm BZ}$ for the Michel outgoing pair $\tau'_\infty=-2\OmF P$: the constraints know which superposition radiates. At finite radius the same constraint reads $\dot M(r)=\tfrac23\OmF P[\OmF P+h\tau']$, so the pure Vaidya form is the asymptotic statement, exact where the wind is free. The $\sin^2\theta$ anisotropy of the wind stress is absorbed by the stationary quadrupole sector through logarithmic tails, whose $\ell=2$ projections are gauge invariant (the $\ell=0$ log coefficients carry residual radial gauge freedom), $V\to-\tfrac89\tau'^2_\infty\ln r$ and $\delta g_{\theta\theta}/g_{\theta\theta}\to\tfrac{16}9\tau'^2_\infty\ln r$.

\subsection{Composite and third order}
The pieces join into a matched solution: outer quasi-stationary odd sector; horizon layer, drained by Equation~\ref{eq:transport} with gravity slaved; even sector with stationary towers plus the unique Vaidya drift of Equation~\ref{eq:mdotgen}; far zone matched to the Michel wind. What remains quantitative is the pair of second-order coefficients in $\OmF/\OmH=\tfrac12[1+\delta_2(p)(M\OmH)^2+\dots]$, with $\delta_2$ the equatorial value, which is also the amplitude of the $\sin^2\theta$ profile of $\delta\OmF$ derived below, and $P_{\rm BZ}=\tfrac16\OmH^2P^2[1+C_2(p)(M\OmH)^2+\dots]$. Their governing system is the third-order odd sector, which passes the vacuum benchmark exactly: all pure-gravity parts vanish on the $O(a^3)$ Kerr metric transformed to the ingoing gauge, itself verified Ricci-flat through third order. Three structural theorems follow. By $t$--$\varphi$ parity the absolute second-order correction to any rotation frequency vanishes, so $\delta_2$ multiplies the relative $(M\OmH)^2$ term, which is third order in the odd sector. The system forces differential rotation: with a uniform shift alone the $\ell=3$ toroidal rows are inconsistent, and closure requires $\OmF(\Psi)=\OmH/2+\epsilon^2[b_0+b_2(1-\Psi/P)^2]$, with the harmonics entering the outgoing-wind matching line by line. And the linear-in-$v$ rows reproduce the spin-down law at the field-equation level, the drift of $g_{v\varphi}$ obeying a drag-type equation whose far tail is $2\dot J/r^3$. Two further results are exact, and together they say what the third-order system does and does not determine.

First, a degeneracy theorem. Differentiating every third-order row with respect to the rotation constants $d_0,d_1,d_2$, which are those of the third-order frequency and are not to be confused with the coefficient $\delta_2(p)$ below, and to the secular drift amplitudes $a_1$, $a_3$ of the toroidal potential, gives, at fixed values of the three combinations of Equation~\ref{eq:invariants},
\begin{equation}
 \frac{\partial(\text{row})}{\partial d_0}\equiv0,
 \qquad
 \frac{\partial(\text{row})}{\partial d_2}\equiv0
 \label{eq:degen}
\end{equation}
identically, in all seven row families. The system depends on those five quantities only through the three combinations
\begin{equation}
 \hat e_0=a_1+P(d_0-d_1),
 \quad
 \hat e_2=a_3+3Pd_2,
 \quad
 \hat d=d_1-d_2 ,
 \label{eq:invariants}
\end{equation}
which are the coefficients of the third-order ingoing electric field, $F_{v\theta}|_3/\sin\theta=\hat e_0+2P\hat d\cos\theta +\hat e_2\cos^2\theta$. Splitting that field into a shift of $\OmF$ and a secular drift of the toroidal potential is a reparametrization of the first-order family, the same $\OmF\leftrightarrow \mathcal C_0$ freedom that Appendix~\ref{app:working} exhibits. What is physical is the field, and the power correction is its $\sin^3\theta$-weighted average,
\begin{equation}
 C_2=\frac{2}{P\OmF}\Big[\hat e_0(\infty)+\tfrac15\hat e_2(\infty)\Big].
 \label{eq:C2inv}
\end{equation}

Second, the probe limit of Equation~\ref{eq:C2inv} is known exactly and the formula reproduces it. Re-expanding the closed-form split-monopole results on Kerr, Blandford and Znajek's $O(a^2)$ power together with Tanabe and Nagataki's $O(a^4)$ term \citep{1977MNRAS.179..433B,2008PhRvD..78b4004T}, in the exact horizon frequency instead of in spin, gives
\begin{equation}
 C_2(0)=\frac{536-48\pi^2}{45}=1.3835331 ,
 \label{eq:C2probe}
\end{equation}
consistent with the fitted $1.38$ \citep{2010ApJ...711...50T} and with the matched-asymptotics coefficient of \citet{2022JCAP...07..032C}. Feeding the probe profile $\OmF(\theta)-\OmH/2=(1-4U_0)(M\OmH)^2\OmH\sin^2\theta$, with $U_0=R(r_+)$ as below, $1-4U_0=(67-6\pi^2)/18$, into Equation~\ref{eq:C2inv} returns $(536-48\pi^2)/45$ algebraically. The assembly is therefore correct, and what remains is the numerical determination of $\hat e_0$ and $\hat e_2$ at finite flux.

That determination requires one further ingredient, and identifying it resolves the problem. The numerical side is settled first. Projecting the angular dependence exactly, with no quadrature, and fixing the one genuine zero mode of the tower system, the second-order towers solve to a median residual of $2\times10^{-11}$ across $p\in[0.05,0.5]$ with no rows discarded, and return the analytic far-field ladder in the $\ln r$ coefficients $u_L$ and $v_L$ of $U$ and $V$ ($u_L/v_L=-2$, $v_L/\tau'^2_\infty=-8/9$, $u_L/\tau'^2_\infty=16/9$) to nine digits. That ladder enters the solve as a soft constraint, so the agreement is a consistency check; the sharper test is the zero mode below, which is imposed nowhere. That zero mode is a reparametrization of the first-order family: a constant rescaling of the azimuthal circumference with the induced renormalization of the flux, $P\to P(1+\epsilon^2\OmF c_\ast)$, and the ratio of its components is exactly $p/2$ to $10^{-22}$, independent of resolution. Improving the residual by seven orders of magnitude did not move the rotation constants at all, which is the sharpest statement that precision was never the missing ingredient.

What was missing is a boundary condition, and the second-order flux obeys an equation that supplies it. Writing the quadrupole flux as $\Psi_2=P\,R(r)\sin^2\theta\cos\theta$, which is the complete angular content at this order, the force-free condition reduces on any slowly rotating background to
\begin{equation}
 r^2\big(hR'\big)'-6R=\tfrac32 Q_2(r)
 +\frac{2r^2}{h}\,\frac{(\OmF-w)^2-I_1^2}{a^2} ,
 \label{eq:stream}
\end{equation}
where $Q_2$ is the $\ell=2$ part of the second-order metric combination $(\alpha^2g_{rr})/(g_{\theta\theta}g_{\varphi\varphi})$, $\alpha$ being the lapse and $I_1$ the first-order current amplitude. The metric enters through that one combination, undifferentiated. The two boundary conditions are forced: the bracket must vanish at the horizon, which is the Znajek condition $I_1=\OmH-\OmF$, and at infinity, which is the Michel condition $I_1=\OmF$. Their conjunction is $\OmF=\OmH/2$, which is the solvability statement quoted in the body. With those conditions imposed, and $R$ analytic at the double indicial root at $r_+$ with no $r^3$ growth at infinity, the problem is a two-point boundary value problem with a unique solution. Neither a least-squares system nor any regularization enters.

The probe limit is then a test, and it passes:
\begin{equation}
 R(2M)=0.14191147787
 =\tfrac{1}{72}(6\pi^2-49) ,
\end{equation}
a relative agreement of $4\times10^{-16}$, at double-precision round-off, with the correct sign, together with the far field $R\to\frac18(r_0/r)$ in units of the probe horizon radius $r_0=2M$, its logarithmic coefficient $1/40$, and its $(r_0/r)^2$ constant $-11/800$, which is exact rather than approximate: the $r^{-2}$ coefficient about $\ln r$ is $-11/200-(\ln2)/10$, and converting to the origin $\ln(r/r_0)$ gives $-11/800$ identically. The rotation correction follows algebraically from the horizon matching, in the form $\delta\Omega(\theta)=\tfrac14\OmH a^2[\gamma_2(\theta)-U_0\sin^2\theta]$ with $U_0=R(r_+)$ and $\gamma_2$ from the horizon metric. Here the degeneracy of Equation~\ref{eq:invariants} reappears in its geometric guise: under a relabeling of the polar angle $\gamma_2$ and $U_0$ shift oppositely and only their difference is physical, which is why a formulation carrying the flux and the metric as independent unknowns cannot fix either. Taking $p\to0$ reproduces $(d_0-d_1)=0.4323540885$, $d_2=-0.1441180295$ and $C_2(0)=1.3835330833$, the exact value of Equation~\ref{eq:C2probe}, to ten digits.

At finite flux the same construction gives the corrections themselves. Solving Equation~\ref{eq:stream} on magnetic Kerr--Newman with its first-order drag replaced by the extracting one,
\begin{equation}
 \delta_2(p)=0.8647082\big(1-0.1517p^2-0.0037p^4+0.0074p^6\big),
 \;\; C_2(p)=1.3835331\big(1-0.5372p^2-0.0037p^4+0.0074p^6\big),
 \label{eq:d2C2}
\end{equation}
both fits good to $2\times10^{-5}$ over $0\le p\le\tfrac12$. The two are locked at $C_2=\tfrac85\delta_2$ in the probe limit, but the ratio drifts with flux, reaching $1.440$ at $p=\tfrac12$. That the two fits share their $p^4$ and $p^6$ coefficients is structural rather than accidental: the two differ only through $R_\infty/2=p^2/3$ on top of a common $\Lambda-\bar U_0$, so $\delta_2/\delta_2(0)-C_2/C_2(0)=0.38549\,p^2$ exactly and every coefficient beyond $p^2$ coincides. Both corrections are built from the same third-order electric field: with the drift amplitudes set to zero, $\hat e_0=P(d_0-d_1)$ and $\hat e_2=3Pd_2$, and the numbers above give $\delta_2=2(d_0-d_1)$ together with $d_2=-\tfrac13(d_0-d_1)$, that is $\hat e_2=-\hat e_0$, to ten digits. Then Equation~\ref{eq:C2inv} returns $C_2=(2/\OmF)(d_0-d_1)(1-\tfrac15)$, and the impedance match $\OmF=\tfrac12\OmH$ turns the remaining $\tfrac45$ into $C_2/\delta_2=\tfrac85$. Part of that has a deeper root than the algebra: axis regularity makes the horizon two-metric conformally round, so $\gamma_2\propto\sin^2\theta$ and hence so is $\delta\OmF$, and this holds exactly at every flux and for either drag profile. The step from that shape to the number $\tfrac85$ is the conditional one, since it also requires $R(\infty)=0$; what the solve returns is the pair $C_2=\tfrac45(\Lambda-\bar U_0)$ and $\delta_2=\tfrac12[\Lambda-(\bar U_0-R_\infty)]$, with $\Lambda=(\sigma+1)^2$, $\sigma=\sqrt{1-p^2}$, and $\bar U_0$, the value of $R(r_+)$, and $R_\infty$ both in the $(M\OmH)^2$ normalization in which $R_\infty=2p^2/3$, so that $\bar U_0=16U_0$ in the probe limit; hence $C_2/\delta_2=\tfrac85(\Lambda-\bar U_0)/[\Lambda-(\bar U_0-R_\infty)]$, which at $p=0$ has $\Lambda=4$, $\bar U_0=2(6\pi^2-49)/9$ and reduces to the two closed forms identically. $C_2$ is independent of $R_\infty$ while $\delta_2$ carries it with $\partial\delta_2/\partial R_\infty=\tfrac12$. Values:
\begin{center}
\begin{tabular}{@{}l@{\ \ \ }c@{\ \ \ }c@{\ \ \ }c@{}}
\hline
$p$ & $\delta_2$ & $C_2$ & $C_2/\delta_2$ \\
\hline
0.001 & 0.8647081 & 1.3835325 & 1.6000 \\
0.01  & 0.8646951 & 1.3834589 & 1.5999 \\
0.05  & 0.8643803 & 1.3816752 & 1.5985 \\
0.1   & 0.8633963 & 1.3761008 & 1.5938 \\
0.2   & 0.8594574 & 1.3537986 & 1.5752 \\
0.3   & 0.8528840 & 1.3166144 & 1.5437 \\
0.5   & 0.8318232 & 1.1975837 & 1.4397 \\
\hline
\end{tabular}
\end{center}
Below $p\simeq0.05$ the quadratic term alone accounts for the entries to the digits shown: at $p=0.01$ the flux costs $1.5\times10^{-5}$ of $\delta_2$ and $5.4\times10^{-5}$ of $C_2$. On this geometry self-gravity reduces both corrections, but not by the same fraction. The power correction stays positive, so the $O((M\OmH)^2)$ term still enhances $P_{\rm BZ}$, and the enhancement is eaten into by $2.1\%$ at $p=0.2$ and by $13.4\%$ at $p=0.5$; the impedance correction is far more robust, falling by only $0.6\%$ and $3.8\%$ at the same two fluxes. The effect becomes appreciable only for a genuinely field-dominated engine. A companion paper carries the rotation expansion to exact coefficients through at least eighth order.

The magnetosphere's own second-order gravitational response is computed rather than estimated. The $\ell=2$ block closes in three rows, the $(\theta\theta)-(\varphi\varphi)$ row being algebraic and carrying the same Znajek--Michel bracket that governs Equation~\ref{eq:stream}, and the construction returns vacuum Kerr and exact magnetically charged Kerr--Newman through $O(a^2)$ to machine zero. At fixed first-order drag, replacing the electrovac $Q_2,\gamma_2$ by the engine's own moves $\delta_2$ by $-0.05\%$ at $p=0.1$, $-0.19\%$ at $p=0.2$ and $+0.18\%$ at $p=0.5$: the second-order metric is worth less than half a percent, and the geometric sensitivity lives in the first-order drag. This calculation covers only the stationary even metric at fixed drag. The full $O(P^2)$ back-reaction also contains the spin-down drift and the odd $O(a^3P^2)$ metric that renormalizes the rotation of the horizon, and its effect is large. The $p^2$ coefficient changes from $-0.743$ in $C_2$ to $-4.36$ or $+5.64$ in the self-consistent $C_2^{\rm sc}$, depending on whether $\OmH$ is measured on the apparent or on the event horizon of the drifting hole (Appendix~\ref{app:ceilingNL}), and the self-consistent counterpart of the equatorial coefficient $\delta_2$ acquires a non-analytic term, $\delta_2^{\rm sc}(p,x)=0.8647+p^2[\ln x+0.11(8)]$ with $x=M\OmH$, from the field angular momentum stored in the wind out to the light cylinder. The fits of Equation~\ref{eq:d2C2} and the table therefore give the coefficients of this geometry; those of the fully self-consistent engine differ. The first-order drag used here is the extracting profile, and the alternative drag profile, the one that keeps $\OmF$ as the source frequency while imposing $w(r_+)=\OmH$, is excluded twice over, since it balances no angular-momentum budget and is the logarithmic branch of the rigidity theorem itself.

One obstruction survives, and it is the Znajek--Michel bracket again. At $r_+$ the $(rr)$ row carries $E^2+B_T^2$ in the electric and toroidal magnetic amplitudes, and the Znajek condition makes $E=B_T$ there, so the bracket is $2E^2\neq0$: the horizon radiates, and since $P_{\rm BZ}$ is itself $O(a^2)$ the second-order metric cannot be stationary either. The far end is settled, and it is what separates the two coefficients. With the extracting drag the stream function tends to $R_\infty=2p^2/3$ rather than to zero, and the matching fixes $\delta\OmF$ through $R(r_+)-R(\infty)$. That limit is not a boundary condition one may impose: the homogeneous solutions at large $r$ are $r^3$ and $r^{-2}$, neither tending to a nonzero constant, so a drag with a $2P^2\OmH/r^2$ tail leaves the constant in the source and the $r^{-3}$ balance forces it. Setting $R(\infty)=0$ while retaining that drag assigns the stream function two values at infinity in one calculation; carrying it consistently gives $\partial\delta_2/\partial R_\infty=\tfrac12$ identically in $p$. The consequence is the split already visible in the table: $C_2$ depends on $R(r_+)$ alone and falls by $13.4\%$ at $p=\tfrac12$, while $\delta_2$ depends on the difference and falls by only $3.8\%$. The electrovac drag, whose $r^{-3}$ tail gives $R_\infty=0$, is the case in which the two fractions coincide and the ratio collapses to $\tfrac85$.

\section{Extremality of the working state, and its throat}\label{app:extremal}

\subsection{The budget: electrovac versus force-free}
The budget of Equation~\ref{eq:budget} is exact for the electrovac member of the family, and the force-free state carries an $O(a^2)$ energy of its own. Whether that moves the degeneracy point is an $\ell=0$ question, and it is local. Write the horizon area as $4\pi R_{\rm H}^2$ and $q=P/R_{\rm H}$. In the areal gauge the $\ell=0$ metric function is $\mathcal F=h-2m(r)/r$, the horizon is its root, and extremality is $\mathcal F'=0$ there, so with $m$ normalized to vanish at the horizon the condition reads $1-q^2=2m'(R_{\rm H})$. The mass function is sourced by the frame dragging and by the second-order electromagnetic energy, and at the horizon the source collapses. Let $\Omega(r)$ be the angular velocity profile carried by the electric field, $\Omega=a/r^2$ for Kerr--Newman and $\Omega=\OmF$ constant for a degenerate one. Both branches have $w(r_+)=\Omega(r_+)$, which kills every term carrying $w-\Omega$, leaving
\begin{equation}
 m'(r_+)=\tfrac1{12}r_+^4w'^2+\tfrac16P^2r_+^2\Omega'^2 ,
 \label{eq:massfn}
\end{equation}
the drag term and the radial electric field. The condition becomes
\begin{equation}
 1-q^2=2X(q)\,(\OmH R_{\rm H})^2 ,
 \qquad
 X=3i_H^2+\tfrac23q^2\,\Theta ,
 \label{eq:Xbudget}
\end{equation}
with $i_H=J_H/(\OmH R_{\rm H}^3)$ the horizon's dimensionless moment of inertia, and $\Theta=1$ for the electrovac branch against $\Theta=0$ for the force-free one, where degeneracy makes $\OmF$ radially constant and so removes the radial electric field. Kerr--Newman has $i_H=(3-q^2)/6$ and therefore $X_{\rm elec}=\tfrac34+\tfrac16q^2+\tfrac1{12}q^4$, which is $1$ at $q=1$: there Equation~\ref{eq:Xbudget} is $1-q^2=2(\OmH R_{\rm H})^2$, exactly $a^2+P^2=M^2$, and imposing that exact boundary on $\tfrac16\OmH^2P^2$ returns $1/48$ at $q^2=\tfrac12$, which is Equation~\ref{eq:ceiling}. Both terms of $X$ are smaller for the engine, one because the force-free condition kills it and the other because the dead-state horizon holds less angular momentum for its rotation rate, and
\begin{center}
\begin{tabular}{@{}l@{\ \ }c@{\ \ }c@{\ \ }c@{\ \ }c@{\ \ }c@{}}
\hline
$p$ & 0.2 & 0.5 & $\sqrt{2/3}$ & 0.9 & $\to1$\\
\hline
$X_{\rm ff}/X_{\rm elec}$ & 0.97 & 0.79 & 0.38 & 0.21 & 0\\
\hline
\end{tabular}
\end{center}
so the budget is relaxed, by a factor $2.6$ in the rotational term at the flux that optimizes the power. How much that buys is not settled at this order. The same $O(\OmH^2)$ truncation applied to Kerr--Newman, where the answer is known, overshoots $1/48$ by $18\%$ and moves the optimum from $q^2=\tfrac12$ to $2\sqrt3-3=0.464$; and as $q\to1$ the coefficient $X_{\rm ff}$ vanishes as $\sigma^{2(\Delta-1)}$ with the horizon moment below, so the leading obstruction disappears and the boundary leaves the slow-rotation domain altogether. What the calculation settles is the direction and the size of the effect, and that Equation~\ref{eq:ceiling} is not endangered from this side.

\subsection{The lifetime output}
The same family fixes what a hole can deliver in total. With $J=aM$, $\dot M=-P_{\rm BZ}$ and $\dot J=2\dot M/\OmH$ the time drops out and $P_{\rm BZ}$ cancels, leaving
\begin{equation}
 \frac{dM}{da}=\frac{M\,a}{2r_+^2+a^2},
 \qquad r_+=M+\sqrt{M^2-a^2-P^2} ,
 \label{eq:spindown-traj}
\end{equation}
so the flux magnitude sets only the timescale and the path feels it through $r_+$ alone. In the probe limit $r_+=M(1+\sqrt{1-a^2/M^2})$ and the integral from an extremal start to $a=0$ gives $M_f/M_0=e^{1/4}/\sqrt2$, a lifetime output of $9.21\%$ within the same leading-order evolution model. Carrying $P$ in $r_+$, starting on the extremality boundary and holding $\Phi$ fixed thereafter, gives $9.20\%$, $8.21\%$, $7.33\%$ and $4.67\%$ at $p_0=P/M_0=0.045$, $0.436$, $1/\sqrt3$ and $\sqrt{2/3}$: the flux that makes the jet possible occupies part of the budget and so denies it spin. The horizon area grows along the same track, and in the probe limit the growth is exact. An extremal Kerr start has $r_+=a=M_0$, so $A_0=4\pi(r_+^2+a^2)=8\pi M_0^2$, while the endpoint $a=0$ is Schwarzschild with $r_+=2M_f$ and $A_f=16\pi M_f^2$; hence
\begin{equation}
 \frac{A_f}{A_0}=2\Big(\frac{M_f}{M_0}\Big)^2=2\cdot\frac{e^{1/2}}{2}=\sqrt e ,
 \label{eq:areagrowth}
\end{equation}
the factor two of the extremal horizon cancelling the $\sqrt2$ of the closed form. Carrying the flux, the extremal start still has $r_+=M_0$ but $a_0^2=M_0^2-P^2$, so $A_0=4\pi(2M_0^2-P^2)$, while the endpoint keeps $P$, with $r_+=M_f+\sqrt{M_f^2-P^2}$ and $A_f=4\pi r_+^2$; at $p=\sqrt{2/3}$ this gives $1.567$. The ceiling of Equation~\ref{eq:ceiling} caps the rate and this caps the integral, and self-gravity lowers both. Since $\dot M=-P_{\rm BZ}$ the whole mass drop leaves as jet, while the equal horizon heat $\int T_H\dot S\,dt$ is paid out of rotation and stays in the hole, so the work done by the torque, $\int\OmH|dJ|$, is twice the mass delivered, $18.4\%$; the whole $29.3\%$ of rotational energy that Christodoulou's bound makes available is used up, $9.2\%$ as jet and $20.1\%$ in raising the irreducible mass from $M_0/\sqrt2$ to $M_f$. Holding the flux at the budget maximum throughout would give $1-e^{-1/4}=22.1\%$, but saturation means $\kappa=0$ and $\dot A=8\pi P_{\rm BZ}/\kappa$ diverges, so the area theorem drives the hole off the boundary at once and the number is an envelope rather than a value. A companion paper applies these rates and this total to gamma-ray bursts, and finds the timescales and the energies consistent with the observed ones.

\begin{figure}[t]
\centering
\includegraphics[width=0.45\linewidth]{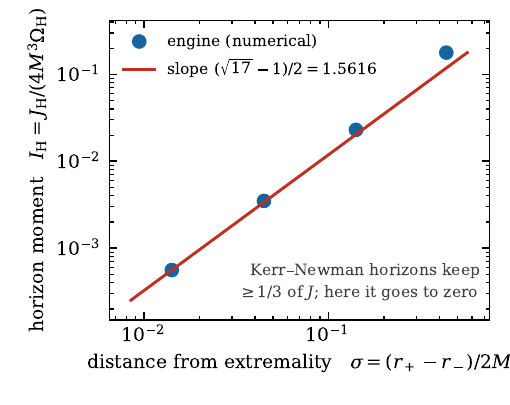}
\caption{The horizon lets go of the spin. Points: the horizon's Komar angular momentum at fixed $\OmH$, computed for the corotating engine as the flux approaches its bound; $\sigma$ measures the remaining distance to extremality. Line: a power law of slope $(\sqrt{17}-1)/2$, the value predicted by the conformal weight of the frame-dragging mode in the Bertotti--Robinson throat, anchored on the last point. Smooth Kerr--Newman fields behave differently: along their extremal family the horizon retains at least $1/3$ of the total angular momentum. Here the horizon moment itself vanishes.}
\label{fig:endpoint}
\end{figure}

\subsection{The ceiling as a circuit}
Because $\kappa_0=1/6\pi$, Equation~\ref{eq:power} is exactly a battery law,
\begin{equation}
 P_{\rm BZ}=\frac16\Big(\frac{\OmH\Phi}{2\pi}\Big)^2\equiv\frac{\mathcal E^2}{6},
 \qquad
 \mathcal E=\OmH P ,
 \label{eq:emf}
\end{equation}
with $\mathcal E$ the electromotive force the spinning hole develops across a hemisphere. A source of internal resistance $\mathcal R$ feeding a matched load delivers $\mathcal E^2/4\mathcal R$, so $\mathcal R=3/2$ and the circuit current is $I_{\rm eff}=\mathcal E/2\mathcal R=\mathcal E/3$; both are pure numbers in geometrized units and therefore independent of the mass. At the optimum $a=M/\sqrt3$, $P=\sqrt{2/3}\,M$, $r_+=M$ and $R_{\rm H}^2=4M^2/3$, so $M\OmH=\sqrt3/4$ and $\mathcal E=\sqrt2/4$, $I_{\rm eff}=\sqrt2/12$, $\mathcal R=3/2$. One geometrized unit of potential is $c^2/\sqrt G=1.043\times10^{27}\,$V, one of current is $c^3/\sqrt G=3.479\times10^{25}\,$A, and one of resistance is $Z_0/4\pi=29.98\,\Omega$, so the optimum reads $3.69\times10^{26}\,$V, $4.10\times10^{24}\,$A, and $44.97\,\Omega$ presented by the source and by the load alike. The circuit closes on itself: the matched terminal voltage $\mathcal E/2$ times the current is $7.56\times10^{50}\,$W, which is $c^5/48G$.

\subsection{The throat}
At $p=1$ the horizon is extremal, $h=(1-M/r)^2$, and the near-horizon limit $r=M(1+\lambda z)$, $t=M\tilde\tau/\lambda$, $\lambda\to0$ gives exactly
\begin{equation}
 ds^2=M^2\Big[-z^2d\tilde\tau^2+\frac{dz^2}{z^2}+d\Omega^2\Big],
 \qquad
 F=M\sin\theta\,d\theta\wedge d\varphi ,
\end{equation}
${\rm AdS}_2\times S^2$ with both radii $M$ and uniform field $B=1/M$: the Bertotti--Robinson universe, which is the massless member of the Kerr--Bertotti--Robinson family \citep{2025PhRvL.135r1401P}. That family's black holes carry a mass parameter and are neither conformally flat nor homogeneous; the throat is the background they are built on, and it is what the Kerr--BR jet program takes as its arena. Here it arrives with an explicit exterior: the disk, the wind, and a far zone that is flat in the weak sense described above. The drag equation becomes $z^2y''=4y$, giving the weight of Equation~\ref{eq:weight}. The electrovac member has instead the first-order drag $w_{\rm KN}=a(2Mr-P^2)/r^4$, analytic at $r=M$ with integer powers of $r-M$ only, so the irrational exponent belongs to the force-free branch. The subdominant exponent $1-\Delta<0$ diverges at $z=0$, so any regular limiting solution, with $w'$ bounded at $z=0$, has $y\sim z^\Delta$ and $w'\sim z^{\Delta-1}\to0$, hence $J_H=-\tfrac16r_+^4w'(r_+)=0$ exactly at $p=1$. Off extremality the gap $\sigma=(r_+-r_-)/2M$ cuts off the throat and the slippage-carrying mode $z^{1-\Delta}$ communicates with the exterior only through $\sigma^{2\Delta-1}=\sigma^{\sqrt{17}}$, while the field sector and the power are untouched at every $p<1$. The share below is a linear-response coefficient: it is computed at first order in rotation, hence for $a\ll M\sigma$, and only then continued to $\sigma\to0$. The two limits do not commute, and the budget $a\le M\sigma$ is what keeps the first of them ahead of the second. Solving the dead-state equation in that order gives the horizon share $I_H\equiv J_H/(4M^3\OmH)$, normalized on its weak-flux value rather than on $R_{\rm H}$ as $i_H$ is, $I_H=0.178,\ 2.3\times10^{-2},\ 3.5\times10^{-3},\ 5.6\times10^{-4}$ at $p=0.9,0.99,0.999,0.9999$, the data of Fig.~\ref{fig:endpoint}. The local logarithmic slope $d\ln I_H/d\ln\sigma$, by central differences on a fine grid, descends through $1.94,1.70,1.61,1.58$ at those four points; pushed to $p=1-10^{-20}$ it reaches $1.5616$, and a three-parameter fit over $\sigma<10^{-3}$ returns the exponent $1.5615528$ against $\Delta-1=1.5615528$, with a residual of $O(\sigma)$. The rotating throat is a different geometry: at zero Kerr--BR mass parameter that family has constant $R_{ab}R^{ab}$, while for extremal magnetic Kerr--Newman $R_{ab}R^{ab}=4P^4/(r^2+a^2\cos^2\theta)^4$, so on the horizon the invariant varies from pole to equator by $(1+a^2/M^2)^4$, a factor $3.42$ at $a/M=0.6$. The throat is warped at every spin.

\section{The ceiling beyond leading order in rotation}\label{app:ceilingNL}

\subsection{Why a finite-spin solution, and on which geometry}
Equation~\ref{eq:ceiling} maximizes the leading term $\tfrac16\OmH^2P^2$ on the budget boundary, where $M\OmH=\sqrt3/4$ is not small. The rotation series does not reach that point: its leading term is accurate to one percent only up to $x=M\OmH\simeq0.08$ ($a\simeq0.3M$), and high-order truncations lose accuracy near the boundary (the companion paper on the rotation expansion gives the full comparison). We therefore solve for the force-free field at finite spin, with no expansion in rotation.

The geometry is the one that produces $C_2(p)$ in Appendix~\ref{app:second}: magnetically charged Kerr--Newman with parameters $(M,a,P)$, $\Delta_{\rm KN}=r^2-2Mr+a^2+P^2$, $\Sigma=r^2+a^2\cos^2\theta$ and $\mathcal A=(r^2+a^2)^2-a^2\Delta_{\rm KN}\sin^2\theta$. It is the rotating continuation of the split-monopole geometry. The Maxwell stress is quadratic in the field and the metric depends on the charge only through $P^2$, so two copies of magnetic Kerr--Newman, with charge $+P$ above the equator and $-P$ below, glue into a spacetime whose field reverses across the equator as a split monopole does, with flux $\Phi=2\pi P$ through each horizon hemisphere. The equator stays totally geodesic, so the sheet is weightless at every spin, and at $a=0$ the construction is the static solution of Appendix~\ref{app:static}. The flux thus enters $\Delta_{\rm KN}$, $r_+$ and the budget as a parameter of the geometry, not as a test field.

This electrovac geometry and the force-free engine differ at $O(a)$ through the frame dragging (Section~\ref{sec:spacetime}), so we replace its first-order drag by the extracting one, $w=w_{\rm KN}+\OmH D(\varrho)$, with $\varrho=r_+/r$ and $D=w_{\rm EX}-w_{\rm KN}$ the difference between the extracting and the first-order Kerr--Newman drag at the same flux, both normalized by $\OmH$. What the fixed geometry leaves out is the weight of the force-free field beyond first order in rotation; Appendix~\ref{app:ceilingBR} estimates its effect at the optimum.

\subsection{The force-free solution}
On this background the stationary axisymmetric force-free field is a flux function $\Psi(r,\theta)$ that solves the stream equation. Its two free functions, the field-line angular velocity and the poloidal current, depend on the flux label $\vartheta=\arccos(1-\Psi/P)$ and are expanded as
\begin{equation}
 \OmF=\OmH\,\varpi(\vartheta),\quad \varpi=\sum_{k<N/2}\varpi_kT_{2k}(\chi),\qquad
 I\propto\sin^2\vartheta\sum_{k<N/2}\mathcal I_kT_{2k}(\chi),
\end{equation}
with $\chi=1-2\vartheta/\pi$, so that both are even about the equator. The coefficient of $\partial_r^2\Psi$ is the light-surface function
\begin{equation}
 \mathcal K=\frac{\Sigma^2\Delta_{\rm KN}-\sin^2\theta\,\Xi^2}{\mathcal A},\qquad
 \Xi=\mathcal A\OmH(\varpi-D)-a(2Mr-P^2),
 \label{eq:Klight}
\end{equation}
which vanishes on an inner and an outer light surface. On each surface the stream equation degenerates to a first-order condition. Imposing it there, with the surface located by bisection at every Newton step, gives exactly the conditions that fix $\varpi_k$ and $\mathcal I_k$; they replace the Znajek and Michel conditions of the rotation expansion.

The unknown $\Psi_1=\Psi-P(1-\cos\theta)$ is represented on fourteen Chebyshev subdomains in $\tilde r=\ln(r/r_+)\in[0,80]$, with $n_t$ nodes per subdomain and edges clustered about the light cylinder $\tilde r\simeq\ln(4/\bar a)$, $\bar a=2r_+\OmH$, and on $N$ Gauss nodes in $\theta$. Both $\Psi_1$ and $\partial_{\tilde r}\Psi_1$ are continuous across the edges, and the outer condition is $\partial_{\tilde r}\Psi_1=0$. The discretized equations are solved by Newton iteration with a forward-mode automatic-differentiation Jacobian. The power is the Znajek flux through the horizon,
\begin{equation}
 P_{\rm BZ}=\tfrac16\OmH^2P^2\,\hat P,\qquad
 \hat P=3\int_0^{\pi}\varpi(1-\varpi)\Big(\frac{\partial_\theta\Psi}{P}\Big)^2\frac{r_+^2+a^2}{r_+^2+a^2\cos^2\theta}\,\sin\theta\,d\theta ,
 \label{eq:Phat}
\end{equation}
normalized so that $\hat P=1$ for the monopole at $\varpi=\tfrac12$ in the slow-rotation limit.

\subsection{Tests}
At small spin $\hat P$ must reproduce the rotation series of the same geometry, $\hat P=\sum_{n,k}C_{n,k}(p)\,x^n\ln^kx$. Its $C_{2,0}$ is the $C_2(p)$ of Equation~\ref{eq:d2C2}, it carries $\ln x$ terms from $x^5$ on through the matching to the far zone, and its coefficients beyond $x^2$ come from the high-order rotation expansion of the companion paper. At $p=0.1$, $a=0.1M$ ($x=0.025190$) the solver, run in extended precision, gives $\hat P=1.000868663290416784$. The series through $x^{24}$ differs from it by $1.4\times10^{-20}$, and the series through $x^4$ by $1.9\times10^{-8}$, the size of the $x^5$ term. Changing $N$ from 24 to 32 and $n_t$ from 16 to 22 changes $\hat P$ by less than $2.4\times10^{-7}$ in the tested near-extremal cases, which is less than $5\times10^{-9}$ in $P_{\rm BZ}G/c^5$.

\subsection{The approach to the boundary and the optimum}
The boundary has $\kappa=0$ and can be approached but not attained. With $\delta\equiv1-(a^2+P^2)/M^2$ the horizon is $r_+=M(1+\sqrt\delta)$, so we extrapolate in $\sqrt\delta$ at fixed $a$. We compute $\hat P$ at $\delta=2\times10^{-4}$, $5\times10^{-4}$, $10^{-3}$ and $2\times10^{-3}$ for eleven spins between $0.45M$ and $0.70M$, refined down to $\delta=5\times10^{-5}$ for $0.52\le a/M\le0.58$, and extrapolate to $\delta=0$ with a cubic in $\sqrt\delta$. At $a=0.55M$, for example, the six values $0.907860$, $0.891281$, $0.878359$, $0.866015$, $0.859476$ and $0.854719$ at $\delta=2\times10^{-3}$ down to $5\times10^{-5}$ extrapolate to $0.84277$. The boundary values, with $P_{\rm BZ}=\tfrac16x^2p^2\hat P$, $x=Ma/(M^2+a^2)$ and $p^2=1-a^2/M^2$ there, are
\begin{center}
\begin{tabular}{@{}l@{\ \ }c@{\ \ }c@{\ \ }c@{\ \ }c@{\ \ }c@{\ \ }c@{\ \ }c@{}}
\hline
$a/M$ & 0.45 & 0.50 & 0.52 & 0.55 & 0.577 & 0.60 & 0.70\\
\hline
$\hat P$ & 0.8747 & 0.8577 & 0.8515 & 0.8428 & 0.8355 & 0.8298 & 0.8093\\
$48P_{\rm BZ}G/c^5$ & 0.7815 & 0.8234 & 0.8327 & 0.8385 & 0.8355 & 0.8270 & 0.7287\\
\hline
\end{tabular}
\end{center}
A quartic fit of $\hat P(a)$ through all eleven boundary values, maximized with the kinematic factor, gives
\begin{equation}
 P_{\rm BZ}^{\max}=0.8386(2)\,\frac{c^5}{48G}=\frac{c^5}{57.24\,G}=6.3\times10^{57}\,{\rm erg\,s^{-1}} ,
 \label{eq:ceilingNL}
\end{equation}
at $a=0.5532M$ and $P=0.8331M$, $16\%$ below Equation~\ref{eq:ceiling}. The leading-order optimum $a^2=M^2/3$ gives $0.8355$ on the same footing, so moving the optimum gains only $0.4\%$; the reduction comes from the decline of $\hat P$ with spin, which the rotation series already shows as the negative $x^4$ coefficient $C_4$. Extrapolating with polynomials of degree two, three and four in $\sqrt\delta$ changes $48P_{\rm BZ}^{\max}G/c^5$ by $8\times10^{-5}$, and the quoted error, $2\times10^{-4}$, covers this and the resolution changes.

\subsection{Back-reaction at the optimum}\label{app:ceilingBR}
The geometry above responds to the field only through the first-order drag. The self-consistent expansion keeps the full metric back-reaction at relative order $P^2$. In units of the reference mass ($M=1$) it uses the metric $g_{\rm Kerr}(a(v),M(v))+P^2H$ in the ingoing gauge $H_{rr}=H_{r\theta}=H_{v\theta}=0$, with the adiabatic drifts $a(v)=a[1+P^2(\alpha_1+\alpha_3a^2)v]$ and $M(v)=1+P^2m_1a^2v$, where $\alpha_1=-1/12$, $m_1=-1/96$ (the horizon energy flux) and $\alpha_3=-0.0140196$ (the angular-momentum balance), and it imposes the Michel condition at fixed retarded time. The result is
\begin{equation}
 C_2^{\rm sc}(p)=1.3835331+\mu_H\,p^2+O(p^4) ,
 \label{eq:C2sc}
\end{equation}
against $1.3835331-0.7432\,p^2$ from Equation~\ref{eq:d2C2}. The coefficient $\mu_H$ is invariant to twelve digits under all gauge and normalization modes of the second-order metric and independent of the far-zone constants of $\OmF$, but it depends on which horizon defines $x=M\OmH$. Because the hole drifts, its event horizon, $g^{rr}=2\dot r_Hg^{rv}$ with $\dot r_H=P^2a^2(2m_1-\alpha_1)=a^2P^2/16$, and its apparent horizon, $g^{rr}=0$, are distinct surfaces at this order, with different generator angular velocities; the horizon also rotates differentially at $O(a^2P^2)$. Averaging the generator angular velocity over the area gives $\mu_H=-4.3558$ on the apparent horizon and $\mu_H=+5.6442$ on the event horizon, whose generators $\ell^a=g^{ar}-\dot r_Hg^{av}$ rotate at $(g^{r\varphi}-\dot r_Hg^{v\varphi})/(g^{rv}-\dot r_Hg^{vv})$; the polar and equatorial values span $-3.01$ to $-5.03$ and $+6.99$ to $+4.97$.

At fixed $\OmH$ and flux, the power correction relative to the leading term is $(\mu_H+0.7432)\,p^2(M\OmH)^2$, that is $-3.61\,p^2(M\OmH)^2$ (apparent) or $+6.39\,p^2(M\OmH)^2$ (event horizon). At the optimum, $p^2=\tfrac23$ and $(M\OmH)^2=\tfrac3{16}$, this is $-0.45$ or $+0.80$ of the leading term, and its sign depends on which horizon defines $\OmH$. The first-order back-reaction shifts the impedance match: keeping the drift in the $O(a)$ rows, horizon regularity and the outgoing condition give $\OmF=\tfrac12\OmH-aP^2\alpha_1/(2r_+)$, with $P^2\alpha_1=-p^2/12$ the spin-down rate of $a(v)$, so that $\OmF/\OmH=\tfrac12[1+p^2/6+O(p^4)]$ at small flux, where $r_+\to2$. This shift enters the power only through the matching factor $4(\OmF/\OmH)(1-\OmF/\OmH)=1-p^4/36+O(p^6)$, $1.5\%$ at $p=0.85$. The second-order correction is therefore not small at the optimum, and the force-free weight also relaxes the budget on which the optimum sits (Appendix~\ref{app:extremal}). Equation~\ref{eq:ceilingNL} is the ceiling on the fixed geometry; its back-reaction correction is of order unity, and its sign is not determined at this order.

\section{Verification}\label{app:verify}

The principal results of this article have been checked either against an independent benchmark, a limit whose value is known in closed form or an external result from the probe literature, or by an internal consistency check, an identity that must hold exactly. Table~\ref{tab:verif} lists the checks and what they returned. Symbolic entries are exact: the quantity shown simplifies to zero, or the two sides agree as expressions, with no tolerance involved. Numerical entries give the benchmark and the achieved agreement. The symbolic and numerical routines behind them are on \href{https://github.com/YWangScience/An-Exact-Engine-for-Black-Hole-Jets}{GitHub}.

Two aspects of the verification are worth singling out because they test the machinery as a whole. The slow-rotation vacuum Kerr--Newman field, which has the non-isorotating profile $a/r^2$ in place of $\OmF$, satisfies the drag equation, Equation~\ref{eq:drag}, identically; since that field is an exact solution obtained independently of anything here, it exercises the whole first-order reduction at once. At third order the same role is played by the vacuum benchmark: all pure-gravity parts of the third-order rows vanish on the $O(a^3)$ Kerr metric transformed to the ingoing gauge, itself verified Ricci-flat to that order.

\begin{table*}
\caption{Verification of the results. Symbolic checks are exact
identities; numerical checks are quoted against an independent
benchmark. $\Sigma=r^2+a^2\cos^2\theta$.}
\label{tab:verif}
\begin{ruledtabular}
\scriptsize
\begin{tabular}{@{}p{0.235\textwidth}p{0.365\textwidth}p{0.325\textwidth}@{}}
Statement & Independent check & Result \\
\hline
\multicolumn{3}{@{}l}{\textit{The solution (Appendix~\ref{app:static})}}\\
Magnetic RN with split-monopole field & all components of
 $G_{\mu\nu}-8\pi T_{\mu\nu}$, and $\nabla_\nu F^{\mu\nu}$ & vanish
 identically \\
Massless sheet & extrinsic curvature of $\theta=\pi/2$ from both sides &
 $K_{ab}=0$ identically \\
Sheet currents, Equations~\ref{eq:Kphi} and \ref{eq:Kr} & distributional limit,
 and numerical smoothing of width $10^{-5}$ & agree to $10^{-9}$ \\
Michel wind of Equation~\ref{eq:michel} force-free on RN & $F\wedge F$, $dF$,
 $F_{\mu\nu}J^\nu$ for arbitrary $\OmF(\Psi)$ & vanish identically \\
Split Kerr--Newman at all spins & $\partial_\theta g_{ab}$ on the
 equator; hemispheric flux; $|K|^2$ & $0$; $2\pi P$; $P^2/4\pi^2r^4$ \\
Bulk field not force-free at spin & $F_{\mu\nu}\tilde F^{\mu\nu}$ from
 the potential via the Maxwell stress &
 $-8P^2ar\cos\theta(r^2-a^2\cos^2\theta)/\Sigma^4$ \\
$\Phi\le2\pi M$ in physical units & Gaussian conversion
 $\Phi_G=2\pi\sqrt{G}M$ & $3.23\times10^{30}(M/M_\odot)$ G\,cm$^2$;
 $2.64\times10^{30}$ at the leading-order optimum \\
\noalign{\smallskip}
\multicolumn{3}{@{}l}{\textit{First order and rigidity (Appendix~\ref{app:first})}}\\
Drag equation, Equation~\ref{eq:drag} & slow-rotation vacuum Kerr--Newman, an
 independently known solution & satisfied identically \\
Vacuum limit & $P\to0$ & returns $w=2J/r^3$ \\
Frobenius residue and log branch & numerical integration of both branches at $p=0.3$, $0.7$ & log coefficient to $4\times10^{-10}$ \\
Curvature singularity on the log branch & Riemann tensor in the regular
 ingoing chart & twelve components carry $w''$ \\
Moment of inertia $1-\tfrac{10}{9}p^2$ & closed-form perturbation theory
 vs.\ direct solve of the boundary value problem & $-1.111111111$ vs.\
 $-10/9$ \\
\noalign{\smallskip}
\multicolumn{3}{@{}l}{\textit{The working state (Appendix~\ref{app:working})}}\\
Lock $h\tau'=P(w-\OmF)$ & eliminating $(r^4w')'$ between the ingoing rows
 & pointwise identity \\
Surface-gravity drainage & amplitude ratio after $v=40M$ vs.\
 $e^{-\kappa v}$ at $p=0.6$ & $5.1366\times10^{-5}$ vs.\ $5.137\times10^{-5}$ \\
$\kappa$ in the decay law & computed from the Killing vector &
 equals $h'(r_+)/2$ identically \\
Michel branches and the factor of two & exact nonlinear stationary
 force-free system & two roots, $\{0,-2\OmF P\sin\theta\}$ \\
Power, Equation~\ref{eq:power} & rebuilt from the Znajek relation; and as a probe
 on slow Kerr & $\OmH^2P^2/6$ both ways \\
Flux identity $T^r{}_v=-\OmF T^r{}_\varphi$ & arbitrary toroidal
 amplitude and non-constant $\OmF(\theta)$ & exact \\
Rates, Equation~\ref{eq:rates} & $\ell=0$ projection of the constraint vs.\ the
 global integral $\oint(-T^r{}_v)dA$ & agree \\
\noalign{\smallskip}
\multicolumn{3}{@{}l}{\textit{Thermodynamics and second order (Section~\ref{sec:budget}, Appendices~\ref{app:second} and \ref{app:extremal})}}\\
Smarr and first law, Equation~\ref{eq:firstlaw} & differentiation of $M(A,\Phi)$;
 Euler relation & residual identically zero \\
Area theorem & $\dot A$ from exact Kerr--Newman evolution at the rates
 of Equation~\ref{eq:rates} & $\dot A=8\pi P_{\rm BZ}/\kappa>0$ \\
Flywheel $\delta m=\tfrac12[\OmH-w]J$ & the $\ell=0$ row is a total
 derivative & exact at every flux \\
Second-order towers & analytic far-field ladder, imposed as a soft
 constraint and returned consistently; the gauge zero mode is imposed
 nowhere & $-2$, $-8/9$, $16/9$ to nine digits; free zero-mode ratio
 $p/2$ to $10^{-22}$ \\
Degeneracy, Equation~\ref{eq:degen} & differentiate every third-order row &
 $\partial/\partial d_0=\partial/\partial d_2\equiv0$ \\
Asymptotic moment $J_\infty$, Equation~\ref{eq:Jinf} & constancy of the subtracted $J(r)$ from $10^5M$ to $10^7M$ & constant to six digits;
 $J_\infty/J_H=1-0.2310p^2$ \\
Second-order flux at the horizon & probe value $(6\pi^2-49)/72$ & $0.141911477869$, relative $4\times10^{-16}$ \\
Power correction $C_2(0)$ & $(536-48\pi^2)/45$ from the closed-form
 $O(a^2)$ and $O(a^4)$ results & agrees to ten digits \\
\noalign{\smallskip}
\multicolumn{3}{@{}l}{\textit{Extremality and the endpoint (Appendix~\ref{app:extremal}, and Appendix~\ref{app:ceilingNL} for the nonlinear ceiling)}}\\
Ceiling, Equation~\ref{eq:ceiling} & symbolic maximization; $20$-digit root
 finding; scan of the quarter disk & $1/48$ at $a^2/M^2=1/3$ \\
Nonlinear ceiling, Equation~\ref{eq:ceilingNL} & finite-spin force-free solution on the budget boundary, extrapolated in $\sqrt{\delta}$; $a$ scan & $0.8386(2)\,c^5/48G$ at $a=0.553M$; angular and radial resolution change $P_{\rm BZ}G/c^5$ by $<5\times10^{-9}$ \\
Circuit form of the ceiling & $P_{\rm BZ}=(\OmH\Phi/2\pi)^2/6$ at the
 optimum; terminal power against $c^5/48G$ & $3.7\times10^{26}$ V,
 $4.1\times10^{24}$ A, $45\,\Omega$; agree to five digits \\
Horizon area growth & same integration as the lifetime output &
 $\sqrt e$ (probe), $1.567$ at $p=\sqrt{2/3}$ \\
Conformal weight, Equation~\ref{eq:weight} & $\Delta(\Delta-1)=4$ from the
 throat limit, three independent routes & $(1+\sqrt{17})/2$ \\
Exponent $\Delta-1$ & local slope of $I_H(\sigma)$ pushed to
 $p=1-10^{-20}$ & $1.5615528$ vs.\ $1.5615528$ \\
Decoupling exponent $2\Delta-1$ & mode-amplitude ratio in the matching
 window & $4.127$ vs.\ $\sqrt{17}=4.123$ \\
Kerr--Newman contrast & Komar integral, normalization calibrated on Kerr
 & $J_H/J\to1/3$; $0.855$ at $a=0.9M$, $0.476$ at $a=0.5M$ \\
Warping of the rotating throat & $R_{ab}R^{ab}$ from the Maxwell stress &
 $4P^4/\Sigma^4$, ratio $(1+a^2/M^2)^4$ \\
Mass function of Equation~\ref{eq:massfn} and $X_{\rm elec}$ & $\ell=0$ reduction of
 exact Kerr--Newman in the areal gauge; $\mathcal F'(R_{\rm H})$ against the
 perturbative value & agree, residual $O(a^4)$; $X_{\rm elec}$ both ways
 to $6$ digits \\
$\ell=0$ second-order equation & $G^t{}_t$ built from the full Ricci in
 the areal gauge: does the lapse mode drop out, and does the source
 match Equation~\ref{eq:deltam}? & lapse mode absent identically; source agrees on shell, and reproduces exact Kerr--Newman at every radius \\
Budget of Equation~\ref{eq:Xbudget} at $q\to1$ & exact extremal Kerr--Newman
 $a^2+P^2=M^2$, and $1/48$ recovered & $X_{\rm elec}(1)=1$ \\
\noalign{\smallskip}
\multicolumn{3}{@{}l}{\textit{Back-reaction (Appendix~\ref{app:ceilingBR})}}\\
Drift constants $\alpha_1$, $m_1$ & rates of Equation~\ref{eq:rates} at weak flux: $\Gamma_J=p^2/12M+O(p^4/M)$ and $P_{\rm BZ}=a^2P^2/96M^4$ & expansion coefficients $\alpha_1=-1/12$, $m_1=-1/96$ exactly \\
Shift of the impedance match & small-flux limit of $\OmF=\tfrac12\OmH-aP^2\alpha_1/(2r_+)$ & $\tfrac12(1+p^2/6)$; matching factor $1-p^4/36$ \\
Coefficient $\mu_H$ of Equation~\ref{eq:C2sc} & all gauge and normalization modes of the second-order metric; far-zone constants of $\OmF$ & invariant to twelve digits \\
\end{tabular}
\end{ruledtabular}
\end{table*}

\clearpage
\bibliographystyle{astronomycomm}
\makeatletter
\def\@biblabel#1{}
\bibliography{references_prl}

@ARTICLE{2025PhRvL.135r1401P,
       author = {{Podolsk{\'y}}, Ji{\v{r}}{\'\i} and {Ovcharenko}, Hryhorii},
        title = "{Kerr Black Hole in a Uniform Bertotti-Robinson Magnetic Field: An Exact Solution}",
      journal = {\prl},
         year = 2025,
        month = oct,
       volume = {135},
       number = {18},
          eid = {181401},
        pages = {181401},
          doi = {10.1103/rfgv-ybz5},
archivePrefix = {arXiv},
       eprint = {2507.05199},
 primaryClass = {gr-qc},
       adsurl = {https://ui.adsabs.harvard.edu/abs/2025PhRvL.135r1401P}
}

@ARTICLE{2023AnP...53500635C,
       author = {{Carleo}, Amodio and {Lambiase}, Gaetano and {{\"O}vg{\"u}n}, Ali},
        title = "{Non-Linear Electrodynamics in Blandford-Znajek Energy Extraction}",
      journal = {Annalen der Physik},
         year = 2023,
        month = may,
       volume = {535},
       number = {5},
          eid = {2200635},
        pages = {2200635},
          doi = {10.1002/andp.202200635},
archivePrefix = {arXiv},
       eprint = {2210.11162},
 primaryClass = {gr-qc},
       adsurl = {https://ui.adsabs.harvard.edu/abs/2023AnP...53500635C}
}

@ARTICLE{2022PhRvL.129f1302D,
       author = {{Dvali}, Gia and {K{\"u}hnel}, Florian and {Zantedeschi}, Michael},
        title = "{Vortices in Black Holes}",
      journal = {\prl},
         year = 2022,
        month = aug,
       volume = {129},
       number = {6},
          eid = {061302},
        pages = {061302},
          doi = {10.1103/PhysRevLett.129.061302},
archivePrefix = {arXiv},
       eprint = {2112.08354},
 primaryClass = {hep-th},
       adsurl = {https://ui.adsabs.harvard.edu/abs/2022PhRvL.129f1302D}
}

@ARTICLE{2022JCAP...07..032C,
       author = {{Camilloni}, Filippo and {Dias}, Oscar J.~C. and {Grignani}, Gianluca and {Harmark}, Troels and {Oliveri}, Roberto and {Orselli}, Marta and {Placidi}, Andrea and {Santos}, Jorge E.},
        title = "{Blandford-Znajek monopole expansion revisited: novel non-analytic contributions to the power emission}",
      journal = {\jcap},
         year = 2022,
        month = jul,
       volume = {2022},
       number = {7},
          eid = {032},
        pages = {032},
          doi = {10.1088/1475-7516/2022/07/032},
archivePrefix = {arXiv},
       eprint = {2201.11068},
 primaryClass = {gr-qc},
       adsurl = {https://ui.adsabs.harvard.edu/abs/2022JCAP...07..032C}
}

@ARTICLE{2021PTEP.2021i3E03K,
       author = {{Kimura}, Masashi and {Harada}, Tomohiro and {Naruko}, Atsushi and {Toma}, Kenji},
        title = "{Backreaction of mass and angular momentum accretion on black holes: General formulation of metric perturbations and application to the Blandford-Znajek process}",
      journal = {Progress of Theoretical and Experimental Physics},
         year = 2021,
        month = sep,
       volume = {2021},
       number = {9},
          eid = {093E03},
        pages = {093E03},
          doi = {10.1093/ptep/ptab101},
archivePrefix = {arXiv},
       eprint = {2105.05581},
 primaryClass = {gr-qc},
       adsurl = {https://ui.adsabs.harvard.edu/abs/2021PTEP.2021i3E03K}
}

@ARTICLE{2020JCAP...04..009A,
       author = {{Armas}, Jay and {Cai}, Yangyang and {Comp{\`e}re}, Geoffrey and {Garfinkle}, David and {Gralla}, Samuel E.},
        title = "{Consistent Blandford-Znajek expansion}",
      journal = {\jcap},
         year = 2020,
        month = apr,
       volume = {2020},
       number = {4},
          eid = {009},
        pages = {009},
          doi = {10.1088/1475-7516/2020/04/009},
archivePrefix = {arXiv},
       eprint = {2002.01972},
 primaryClass = {astro-ph.HE},
       adsurl = {https://ui.adsabs.harvard.edu/abs/2020JCAP...04..009A}
}

@ARTICLE{2019ApJ...875L...1E,
       author = {{Event Horizon Telescope Collaboration} and {Akiyama}, Kazunori and {Alberdi}, Antxon and {Alef}, Walter and {Asada}, Keiichi and {Azulay}, Rebecca and {Baczko}, Anne-Kathrin and {Ball}, David and {Balokovi{\'c}}, Mislav and {Barrett}, John and {Bintley}, Dan and {Blackburn}, Lindy and {Boland}, Wilfred and {Bouman}, Katherine L. and {Bower}, Geoffrey C. and {Bremer}, Michael and {Brinkerink}, Christiaan D. and {Brissenden}, Roger and {Britzen}, Silke and {Broderick}, Avery E. and {Broguiere}, Dominique and {Bronzwaer}, Thomas and {Byun}, Do-Young and {Carlstrom}, John E. and {Chael}, Andrew and {Chan}, Chi-kwan and {Chatterjee}, Shami and {Chatterjee}, Koushik and {Chen}, Ming-Tang and {Chen}, Yongjun and {Cho}, Ilje and {Christian}, Pierre and {Conway}, John E. and {Cordes}, James M. and {Crew}, Geoffrey B. and {Cui}, Yuzhu and {Davelaar}, Jordy and {De Laurentis}, Mariafelicia and {Deane}, Roger and {Dempsey}, Jessica and {Desvignes}, Gregory and {Dexter}, Jason and {Doeleman}, Sheperd S. and {Eatough}, Ralph P. and {Falcke}, Heino and {Fish}, Vincent L. and {Fomalont}, Ed and {Fraga-Encinas}, Raquel and {Freeman}, William T. and {Friberg}, Per and {Fromm}, Christian M. and {G{\'o}mez}, Jos{\'e} L. and {Galison}, Peter and {Gammie}, Charles F. and {Garc{\'\i}a}, Roberto and {Gentaz}, Olivier and {Georgiev}, Boris and {Goddi}, Ciriaco and {Gold}, Roman and {Gu}, Minfeng and {Gurwell}, Mark and {Hada}, Kazuhiro and {Hecht}, Michael H. and {Hesper}, Ronald and {Ho}, Luis C. and {Ho}, Paul and {Honma}, Mareki and {Huang}, Chih-Wei L. and {Huang}, Lei and {Hughes}, David H. and {Ikeda}, Shiro and {Inoue}, Makoto and {Issaoun}, Sara and {James}, David J. and {Jannuzi}, Buell T. and {Janssen}, Michael and {Jeter}, Britton and {Jiang}, Wu and {Johnson}, Michael D. and {Jorstad}, Svetlana and {Jung}, Taehyun and {Karami}, Mansour and {Karuppusamy}, Ramesh and {Kawashima}, Tomohisa and {Keating}, Garrett K. and {Kettenis}, Mark and {Kim}, Jae-Young and {Kim}, Junhan and {Kim}, Jongsoo and {Kino}, Motoki and {Koay}, Jun Yi and {Koch}, Patrick M. and {Koyama}, Shoko and {Kramer}, Michael and {Kramer}, Carsten and {Krichbaum}, Thomas P. and {Kuo}, Cheng-Yu and {Lauer}, Tod R. and {Lee}, Sang-Sung and {Li}, Yan-Rong and {Li}, Zhiyuan and {Lindqvist}, Michael and {Liu}, Kuo and {Liuzzo}, Elisabetta and {Lo}, Wen-Ping and {Lobanov}, Andrei P. and {Loinard}, Laurent and {Lonsdale}, Colin and {Lu}, Ru-Sen and {MacDonald}, Nicholas R. and {Mao}, Jirong and {Markoff}, Sera and {Marrone}, Daniel P. and {Marscher}, Alan P. and {Mart{\'\i}-Vidal}, Iv{\'a}n and {Matsushita}, Satoki and {Matthews}, Lynn D. and {Medeiros}, Lia and {Menten}, Karl M. and {Mizuno}, Yosuke and {Mizuno}, Izumi and {Moran}, James M. and {Moriyama}, Kotaro and {Moscibrodzka}, Monika and {M{\"u}ller}, Cornelia and {Nagai}, Hiroshi and {Nagar}, Neil M. and {Nakamura}, Masanori and {Narayan}, Ramesh and {Narayanan}, Gopal and {Natarajan}, Iniyan and {Neri}, Roberto and {Ni}, Chunchong and {Noutsos}, Aristeidis and {Okino}, Hiroki and {Olivares}, H{\'e}ctor and {Ortiz-Le{\'o}n}, Gisela N. and {Oyama}, Tomoaki and {{\"O}zel}, Feryal and {Palumbo}, Daniel C.~M. and {Patel}, Nimesh and {Pen}, Ue-Li and {Pesce}, Dominic W. and {Pi{\'e}tu}, Vincent and {Plambeck}, Richard and {PopStefanija}, Aleksandar and {Porth}, Oliver and {Prather}, Ben and {Preciado-L{\'o}pez}, Jorge A. and {Psaltis}, Dimitrios and {Pu}, Hung-Yi and {Ramakrishnan}, Venkatessh and {Rao}, Ramprasad and {Rawlings}, Mark G. and {Raymond}, Alexander W. and {Rezzolla}, Luciano and {Ripperda}, Bart and {Roelofs}, Freek and {Rogers}, Alan and {Ros}, Eduardo and {Rose}, Mel and {Roshanineshat}, Arash and {Rottmann}, Helge and {Roy}, Alan L. and {Ruszczyk}, Chet and {Ryan}, Benjamin R. and {Rygl}, Kazi L.~J. and {S{\'a}nchez}, Salvador and {S{\'a}nchez-Arguelles}, David and {Sasada}, Mahito and {Savolainen}, Tuomas and {Schloerb}, F. Peter and {Schuster}, Karl-Friedrich and {Shao}, Lijing and {Shen}, Zhiqiang and {Small}, Des and {Sohn}, Bong Won and {SooHoo}, Jason and {Tazaki}, Fumie and {Tiede}, Paul and {Tilanus}, Remo P.~J. and {Titus}, Michael and {Toma}, Kenji and {Torne}, Pablo and {Trent}, Tyler and {Trippe}, Sascha and {Tsuda}, Shuichiro and {van Bemmel}, Ilse and {van Langevelde}, Huib Jan and {van Rossum}, Daniel R. and {Wagner}, Jan and {Wardle}, John and {Weintroub}, Jonathan and {Wex}, Norbert and {Wharton}, Robert and {Wielgus}, Maciek and {Wong}, George N. and {Wu}, Qingwen and {Young}, Ken and {Young}, Andr{\'e}},
        title = "{First M87 Event Horizon Telescope Results. I. The Shadow of the Supermassive Black Hole}",
      journal = {\apjl},
         year = 2019,
        month = apr,
       volume = {875},
       number = {1},
          eid = {L1},
        pages = {L1},
          doi = {10.3847/2041-8213/ab0ec7},
archivePrefix = {arXiv},
       eprint = {1906.11238},
 primaryClass = {astro-ph.GA},
       adsurl = {https://ui.adsabs.harvard.edu/abs/2019ApJ...875L...1E}
}

@ARTICLE{2018PhRvD..98h4056G,
       author = {{Grignani}, Gianluca and {Harmark}, Troels and {Orselli}, Marta},
        title = "{Existence of the Blandford-Znajek monopole for a slowly rotating Kerr black hole}",
      journal = {\prd},
         year = 2018,
        month = oct,
       volume = {98},
       number = {8},
          eid = {084056},
        pages = {084056},
          doi = {10.1103/PhysRevD.98.084056},
archivePrefix = {arXiv},
       eprint = {1804.05846},
 primaryClass = {gr-qc},
       adsurl = {https://ui.adsabs.harvard.edu/abs/2018PhRvD..98h4056G}
}

@ARTICLE{2015PhRvD..91f4067P,
       author = {{Pan}, Zhen and {Yu}, Cong},
        title = "{Fourth-order split monopole perturbation solutions to the Blandford-Znajek mechanism}",
      journal = {\prd},
         year = 2015,
        month = mar,
       volume = {91},
       number = {6},
          eid = {064067},
        pages = {064067},
          doi = {10.1103/PhysRevD.91.064067},
archivePrefix = {arXiv},
       eprint = {1503.05248},
 primaryClass = {astro-ph.HE},
       adsurl = {https://ui.adsabs.harvard.edu/abs/2015PhRvD..91f4067P}
}

@ARTICLE{2014MNRAS.445.2500G,
       author = {{Gralla}, Samuel E. and {Jacobson}, Ted},
        title = "{Spacetime approach to force-free magnetospheres}",
      journal = {\mnras},
         year = 2014,
        month = dec,
       volume = {445},
       number = {3},
        pages = {2500-2534},
          doi = {10.1093/mnras/stu1690},
archivePrefix = {arXiv},
       eprint = {1401.6159},
 primaryClass = {astro-ph.HE},
       adsurl = {https://ui.adsabs.harvard.edu/abs/2014MNRAS.445.2500G}
}

@ARTICLE{2014PhRvD..89b4041L,
       author = {{Lasota}, J.-P. and {Gourgoulhon}, E. and {Abramowicz}, M. and {Tchekhovskoy}, A. and {Narayan}, R.},
        title = "{Extracting black-hole rotational energy: The generalized Penrose process}",
      journal = {\prd},
         year = 2014,
        month = jan,
       volume = {89},
       number = {2},
          eid = {024041},
        pages = {024041},
          doi = {10.1103/PhysRevD.89.024041},
archivePrefix = {arXiv},
       eprint = {1310.7499},
 primaryClass = {gr-qc},
       adsurl = {https://ui.adsabs.harvard.edu/abs/2014PhRvD..89b4041L}
}

@ARTICLE{2012LRR....15....7C,
       author = {{Chru{\'s}ciel}, Piotr T. and {Costa}, Jo{\~a}o Lopes and {Heusler}, Markus},
        title = "{Stationary Black Holes: Uniqueness and Beyond}",
      journal = {Living Reviews in Relativity},
         year = 2012,
        month = dec,
       volume = {15},
       number = {1},
          eid = {7},
        pages = {7},
          doi = {10.12942/lrr-2012-7},
archivePrefix = {arXiv},
       eprint = {1205.6112},
 primaryClass = {gr-qc},
       adsurl = {https://ui.adsabs.harvard.edu/abs/2012LRR....15....7C}
}

@ARTICLE{2011MNRAS.418L..79T,
       author = {{Tchekhovskoy}, Alexander and {Narayan}, Ramesh and {McKinney}, Jonathan C.},
        title = "{Efficient generation of jets from magnetically arrested accretion on a rapidly spinning black hole}",
      journal = {\mnras},
         year = 2011,
        month = nov,
       volume = {418},
       number = {1},
        pages = {L79-L83},
          doi = {10.1111/j.1745-3933.2011.01147.x},
archivePrefix = {arXiv},
       eprint = {1108.0412},
 primaryClass = {astro-ph.HE},
       adsurl = {https://ui.adsabs.harvard.edu/abs/2011MNRAS.418L..79T}
}

@ARTICLE{2010ApJ...711...50T,
       author = {{Tchekhovskoy}, Alexander and {Narayan}, Ramesh and {McKinney}, Jonathan C.},
        title = "{Black Hole Spin and The Radio Loud/Quiet Dichotomy of Active Galactic Nuclei}",
      journal = {\apj},
         year = 2010,
        month = mar,
       volume = {711},
       number = {1},
        pages = {50-63},
          doi = {10.1088/0004-637X/711/1/50},
archivePrefix = {arXiv},
       eprint = {0911.2228},
 primaryClass = {astro-ph.HE},
       adsurl = {https://ui.adsabs.harvard.edu/abs/2010ApJ...711...50T}
}

@ARTICLE{2008PhRvD..78b4004T,
       author = {{Tanabe}, Kentarou and {Nagataki}, Shigehiro},
        title = "{Extended monopole solution of the Blandford-Znajek mechanism: Higher order terms for a Kerr parameter}",
      journal = {\prd},
         year = 2008,
        month = jul,
       volume = {78},
       number = {2},
          eid = {024004},
        pages = {024004},
          doi = {10.1103/PhysRevD.78.024004},
archivePrefix = {arXiv},
       eprint = {0802.0908},
 primaryClass = {astro-ph},
       adsurl = {https://ui.adsabs.harvard.edu/abs/2008PhRvD..78b4004T}
}

@ARTICLE{2004MNRAS.350..427K,
       author = {{Komissarov}, S.~S.},
        title = "{Electrodynamics of black hole magnetospheres}",
      journal = {\mnras},
         year = 2004,
        month = may,
       volume = {350},
       number = {2},
        pages = {427-448},
          doi = {10.1111/j.1365-2966.2004.07598.x},
       adsurl = {https://ui.adsabs.harvard.edu/abs/2004MNRAS.350..427K}
}

@ARTICLE{2003PASJ...55L..69N,
       author = {{Narayan}, Ramesh and {Igumenshchev}, Igor V. and {Abramowicz}, Marek A.},
        title = "{Magnetically Arrested Disk: an Energetically Efficient Accretion Flow}",
      journal = {\pasj},
         year = 2003,
        month = dec,
       volume = {55},
        pages = {L69-L72},
          doi = {10.1093/pasj/55.6.L69},
archivePrefix = {arXiv},
       eprint = {astro-ph/0305029},
 primaryClass = {astro-ph},
       adsurl = {https://ui.adsabs.harvard.edu/abs/2003PASJ...55L..69N}
}

@ARTICLE{1990ApJ...350..518P,
       author = {{Punsly}, Brian and {Coroniti}, Ferdinand V.},
        title = "{Relativistic Winds from Pulsar and Black Hole Magnetospheres}",
      journal = {\apj},
         year = 1990,
        month = feb,
       volume = {350},
        pages = {518},
          doi = {10.1086/168408},
       adsurl = {https://ui.adsabs.harvard.edu/abs/1990ApJ...350..518P}
}

@BOOK{1986bhmp.book.....T,
       author = {{Thorne}, Kip S. and {Price}, Richard H. and {MacDonald}, Douglas A.},
        title = "{Black holes: The membrane paradigm}",
    publisher = {Yale University Press},
         year = 1986,
       adsurl = {https://ui.adsabs.harvard.edu/abs/1986bhmp.book.....T}
}

@ARTICLE{1982MNRAS.198..345M,
       author = {{MacDonald}, D. and {Thorne}, K.~S.},
        title = "{Black-hole electrodynamics - an absolute-space/universal-time formulation}",
      journal = {\mnras},
         year = 1982,
        month = jan,
       volume = {198},
        pages = {345-382},
          doi = {10.1093/mnras/198.2.345},
       adsurl = {https://ui.adsabs.harvard.edu/abs/1982MNRAS.198..345M}
}

@ARTICLE{1978PhRvD..18.3598D,
       author = {{Damour}, Thibaut},
        title = "{Black-hole eddy currents}",
      journal = {\prd},
         year = 1978,
        month = nov,
       volume = {18},
       number = {10},
        pages = {3598-3604},
          doi = {10.1103/PhysRevD.18.3598},
       adsurl = {https://ui.adsabs.harvard.edu/abs/1978PhRvD..18.3598D}
}

@ARTICLE{1977MNRAS.179..457Z,
       author = {{Znajek}, R.~L.},
        title = "{Black hole electrodynamics and the Carter tetrad.}",
      journal = {\mnras},
         year = 1977,
        month = may,
       volume = {179},
        pages = {457-472},
          doi = {10.1093/mnras/179.3.457},
       adsurl = {https://ui.adsabs.harvard.edu/abs/1977MNRAS.179..457Z}
}

@ARTICLE{1977MNRAS.179..433B,
       author = {{Blandford}, R.~D. and {Znajek}, R.~L.},
        title = "{Electromagnetic extraction of energy from Kerr black holes.}",
      journal = {\mnras},
         year = 1977,
        month = may,
       volume = {179},
        pages = {433-456},
          doi = {10.1093/mnras/179.3.433},
       adsurl = {https://ui.adsabs.harvard.edu/abs/1977MNRAS.179..433B}
}

@ARTICLE{1975PhRvD..12.2959R,
       author = {{Ruffini}, R. and {Wilson}, J.~R.},
        title = "{Relativistic magnetohydrodynamical effects of plasma accreting into a black hole}",
      journal = {\prd},
         year = 1975,
        month = nov,
       volume = {12},
       number = {10},
        pages = {2959-2962},
          doi = {10.1103/PhysRevD.12.2959},
       adsurl = {https://ui.adsabs.harvard.edu/abs/1975PhRvD..12.2959R}
}

@ARTICLE{1974ApJ...193..443T,
       author = {{Teukolsky}, S.~A. and {Press}, W.~H.},
        title = "{Perturbations of a rotating black hole. III. Interaction of the hole with gravitational and electromagnetic radiation.}",
      journal = {\apj},
         year = 1974,
        month = oct,
       volume = {193},
        pages = {443-461},
          doi = {10.1086/153180},
       adsurl = {https://ui.adsabs.harvard.edu/abs/1974ApJ...193..443T}
}

@ARTICLE{1974PhRvD..10.1680W,
       author = {{Wald}, Robert M.},
        title = "{Black hole in a uniform magnetic field}",
      journal = {\prd},
         year = 1974,
        month = sep,
       volume = {10},
       number = {6},
        pages = {1680-1685},
          doi = {10.1103/PhysRevD.10.1680},
       adsurl = {https://ui.adsabs.harvard.edu/abs/1974PhRvD..10.1680W}
}

@ARTICLE{1973CMaPh..31..161B,
       author = {{Bardeen}, J.~M. and {Carter}, B. and {Hawking}, S.~W.},
        title = "{The four laws of black hole mechanics}",
      journal = {Communications in Mathematical Physics},
         year = 1973,
        month = jun,
       volume = {31},
       number = {2},
        pages = {161-170},
          doi = {10.1007/BF01645742},
       adsurl = {https://ui.adsabs.harvard.edu/abs/1973CMaPh..31..161B}
}

@ARTICLE{1973ApJ...180L.133M,
       author = {{Michel}, F. Curtis},
        title = "{Rotating Magnetospheres: an Exact 3-D Solution}",
      journal = {\apjl},
         year = 1973,
        month = mar,
       volume = {180},
        pages = {L133},
          doi = {10.1086/181169},
       adsurl = {https://ui.adsabs.harvard.edu/abs/1973ApJ...180L.133M}
}

@ARTICLE{1970PhRvL..25.1596C,
       author = {{Christodoulou}, Demetrios},
        title = "{Reversible and Irreversible Transformations in Black-Hole Physics}",
      journal = {\prl},
         year = 1970,
        month = nov,
       volume = {25},
       number = {22},
        pages = {1596-1597},
          doi = {10.1103/PhysRevLett.25.1596},
       adsurl = {https://ui.adsabs.harvard.edu/abs/1970PhRvL..25.1596C}
}

@ARTICLE{1969Natur.223..690L,
       author = {{Lynden-Bell}, D.},
        title = "{Galactic Nuclei as Collapsed Old Quasars}",
      journal = {\nat},
         year = 1969,
        month = aug,
       volume = {223},
       number = {5207},
        pages = {690-694},
          doi = {10.1038/223690a0},
       adsurl = {https://ui.adsabs.harvard.edu/abs/1969Natur.223..690L}
}

@ARTICLE{1969ApJ...157..869G,
       author = {{Goldreich}, Peter and {Julian}, William H.},
        title = "{Pulsar Electrodynamics}",
      journal = {\apj},
         year = 1969,
        month = aug,
       volume = {157},
        pages = {869},
          doi = {10.1086/150119},
       adsurl = {https://ui.adsabs.harvard.edu/abs/1969ApJ...157..869G}
}

@ARTICLE{1969NCimR...1..252P,
       author = {{Penrose}, Roger},
        title = "{Gravitational Collapse: the Role of General Relativity}",
      journal = {Nuovo Cimento Rivista Serie},
         year = 1969,
        month = jan,
       volume = {1},
        pages = {252},
       adsurl = {https://ui.adsabs.harvard.edu/abs/1969NCimR...1..252P}
}

@ARTICLE{1969JMP....10...70C,
       author = {{Carter}, Brandon},
        title = "{Killing Horizons and Orthogonally Transitive Groups in Space-Time}",
      journal = {Journal of Mathematical Physics},
         year = 1969,
        month = jan,
       volume = {10},
        pages = {70-81},
          doi = {10.1063/1.1664763},
       adsurl = {https://ui.adsabs.harvard.edu/abs/1969JMP....10...70C}
}

\end{document}